# TL-RL-FusionNet: An Adaptive and Efficient Reinforcement Learning–Driven Transfer Learning Framework for Detecting Evolving Ransomware Threats

**Jannatul Ferdous[1], Rafiqul Islam[2], Arash Mahboubi[3] and Md Zahidul Islam[4]**

[1]School of Computing, Mathematics and Engineering, Charles Sturt University, Wagga Wagga, NSW- 2650, Australia
[2]School of Computing, Mathematics and Engineering, Charles Sturt University, Albury, NSW- 2640, Australia
[3]School of Computing, Mathematics and Engineering, Charles Sturt University, Port Macquarie, NSW 2444, Australia
[4]School of Computing, Mathematics and Engineering, Charles Sturt University, Bathurst, NSW 2795, Australia

Corresponding author: Jannatul Ferdous (email: jferdous@csu.edu.au)

**Abstract**
Modern ransomware demonstrates polymorphic and evasive behaviors by frequently modifying its execution patterns to avoid detection. This dynamic nature disrupts feature spaces and hinders the efficacy of static or predefined models. To address this challenge, we present TL-RL-FusionNet, a reinforcement learning (RL)-guided hybrid framework that integrates frozen dual transfer learning (TL) backbones as feature extractors with a lightweight residual multilayer perceptron (MLP) classifier. The RL agent supervises training by adaptively reweighting samples in response to modifications in the observable behaviors of the ransomware. By providing feedback in the form of rewards or penalties, the RL agent emphasizes complex or deceptive cases such as stealthy ransomware or polymorphic variants employing obfuscation while down weighting trivial samples such as benign applications with only simple file I/O operations or ransomware with obvious encryption routines that are consistently classified correctly. This adaptive mechanism enables the model to dynamically adjust its strategy, improving its resilience against evolving ransomware behaviors while maintaining effective classification performance. The framework employs dynamic behavioral features, such as file system operations, registry modifications, network traffic, API calls, and anti-analysis checks, which are extracted from sandbox-generated JSON reports. These features were then transformed into RGB images and processed through frozen EfficientNetB0 and InceptionV3 layers, thereby capturing comprehensive feature representations while ensuring efficiency. The final classification was performed using a lightweight residual MLP supervised by an RL (Q learning) agent. On a balanced dataset of 1,000 samples (500 ransomware, 500 benign), TL-RL-FusionNet achieved 99.1% accuracy, 98.6% precision, 99.6% recall, and 99.74% AUC, outperforming non-RL baselines by up to 2.5% in accuracy and 3.1% in recall. Efficiency profiling further revealed 55% lower training time and 59% reduced RAM usage, confirming that reinforcement learning not only enhances accuracy but also streamlines computation for real-world ransomware detection.

## 1. Introduction

Ransomware has emerged as one of the most destructive forms of malware, causing substantial economic losses, disrupting critical infrastructure, and threatening data integrity. Unlike traditional malware, ransomware encrypts victims' data and demands a ransom, often in cryptocurrency, for the decryption key [1]. Modern ransomware variants employ polymorphism, obfuscation, and anti-analysis strategies, making their detection using conventional approaches increasingly difficult. Additionally, new attack strategies, such as double extortion, in which data are exfiltrated and threatened with public release, have further magnified the threat. The proliferation of Ransomware-as-a-Service (RaaS) [2] has lowered the entry barriers for cybercriminals, and the widespread use of cryptocurrencies facilitates anonymous ransom payments, complicating law-enforcement efforts [3]. As attackers continue to innovate, the need for advanced and adaptive ransomware detection mechanisms has become increasingly urgent.

Ransomware detection typically relies on static and dynamic analyses to identify malicious behaviors. Static analysis inspects the code structure without execution, thereby providing safer and faster detection. However, it faces challenges when dealing with obfuscated or polymorphic ransomware that employs tactics such as packing and encryption techniques. In contrast, dynamic analysis observes program behavior during execution, enabling the detection of hidden or runtime-triggered actions in the program. However, it is resource-intensive, slow, and susceptible to anti-analysis evasion techniques [1].

Early ransomware detection systems employed traditional machine learning (ML) techniques such as Random Forests (RF), Support Vector Machines (SVM), and Logistic Regression (LR). This refers to algorithms that learn from features extracted from static or dynamic analyses to classify benign and malicious activities. However, Traditional ML methods require manual feature engineering, making them labour-intensive and prone to bias. These models cannot learn abstract hierarchies from raw data, limiting their adaptability to evolving ransomware [4].

To overcome the reliance on manual features, researchers have turned to deep learning (DL) methods for ransomware detection because of their ability to extract high-level semantic features from raw data and pattern recognition [5], [6], [7], [8]. DL, a subfield of ML, trains computers for automated prediction by employing multilayer artificial neural networks inspired by human brain neurons. These networks consist of interconnected layers that transform input data using nonlinear activation functions (e.g., ReLU, Sigmoid, Tanh), thereby enabling hierarchical feature extraction from raw input data. The model parameters were optimized through backpropagation, where gradients were propagated backward to iteratively update the network weights and minimize errors. By combining nonlinear activation functions with backpropagation-driven optimization, these models achieve superior performance in the analysis of high-dimensional data. Convolutional Neural Networks (CNNs) [9], [10], [11] and Long Short-Term Memory (LSTM) networks [9] have demonstrated potential in enhancing detection accuracy. Additionally, DL techniques have been widely applied in image-based detection pipelines [11], [12], [9], [13]. However, these methods require large, balanced datasets for effective training and incur high computational costs, making them less practical for lightweight and real-world applications.

Transfer learning (TL) has emerged as a promising solution to these challenges in recent years. TL leverages knowledge from pretrained CNNs (e.g., ImageNet backbones) to reduce training costs and improve generalization on small datasets. TL is a machine learning technique that allows a model to use previously learned knowledge acquired from one problem domain and apply it to another related domain. By reusing pretrained extractors, TL enhances detection precision, speeds training, and enables rapid adaptation to new threats. Several studies have employed TL for malware classification, showing strong generalization with limited data. TL can be applied in two modes: (i) using pretrained models as frozen feature extractors [13], [14], [15], [16] or (ii) full end-to-end fine-tuning CNNs to learn task-specific features [17]. Despite these advantages, TL-based ransomware detection still lacks mechanisms for dynamic adaptation to sample weighting and struggles with high resource demands, which are unsuitable for resource-constrained environments. In our previous work [18], we achieved state-of-the-art accuracy of up to 99.96% by employing full end-to-end fine-tuning of multiple CNNs (ResNet, EfficientNetB0, InceptionV3, Xception, and VGG) models. Although highly accurate, this approach incurs substantial training time and memory costs, limiting its practicality in real time or resource-constrained environments.

To address adaptability gaps, researchers have recently explored the use of RL in cybersecurity applications [19], [20], [21]. RL allows an agent to interact with its environment and adjust its actions based on rewards, making it suitable for dynamic sample reweighting. RL-based strategies can emphasize harder or misclassified samples, reduce overfitting to trivial cases, and improve generalization to previously unseen threats. In ransomware detection, RL offers four key benefits: (i) adaptive learning that prioritizes high-value security samples, (ii) improved efficiency by focusing on computation in critical cases, (iii) enhanced robustness to evolving attack patterns, and (iv) real-time adaptability to changes in the threat landscape. RL also helps address class imbalance in security datasets but lacks CNN-based feature extractors' representational power, necessitating a hybrid framework.

Despite advances in DL and TL for ransomware detection, existing pipelines remain largely non-adaptive, treating all training samples equally. This uniform treatment leaves models vulnerable when facing an adaptive adversary such as ransomware, which frequently alters its execution strategies to evade detection. In such adversarial settings, rare or stealthy behaviors are easily overshadowed by more common patterns [22], reducing the robustness of static or TL-based approaches. RL offers a promising solution as it adaptively adjusts the training focus based on feedback, enabling the model to learn from the adversary's evolving strategies. However, prior applications of RL in security have mostly been restricted to static PE features, Android permissions, or forensic classification tasks.

To overcome these challenges, we developed TL-RL-FusionNet, a novel hybrid architecture that integrates TL-based feature fusion and RL-guided classification with MLP. The RL agent dynamically reweights the training samples by assigning higher weights to misclassified or challenging instances and lower weights to consistently correct instances. This adaptive weighting reduces overfitting to trivial patterns and improves robustness against evasive or difficult-to-detect ransomware behaviors. In this approach, behavioral JSON logs are transformed into structured vectors and then RGB images to leverage pretrained CNNs, which are optimized for image classification and can extract high-level spatial correlations. These images were passed through the pretrained CNN backbones, EfficientNetB0 and InceptionV3, for feature extraction. EfficientNetB0 provides efficient detail capture, whereas InceptionV3 enables multi-scale understanding. Their strengths yield a rich combined representation. The fused deep features were classified using a proposed residual MLP, and a set of baseline models, under the supervision of a Q-learning agent that dynamically adjusted the sample weights during training based on real-time feedback. This hybrid approach ensures a strong defense against ransomware while remaining deployable in resource-constrained environments.

The key contributions of this study are as follows:
1. We proposed a novel hybrid architecture (TL-RL-FusionNet) where a tabular Q-learning agent reweights training samples by assigning higher importance to challenging instances while down-weighting trivial ones. The agent updates Q-values based on prediction outcomes and is re-initialized per fold under stratified 5-fold cross-validation. This adaptive strategy mitigates overfitting, reduces false negatives, and improves accuracy compared to static training.
2. We used two pretrained CNN models to obtain a richer and more balanced feature representation. By keeping both networks frozen, we maintain inference efficiency by avoiding backpropagation through millions of parameters and eliminating the need to store deep activation tensors. Global-average-pooled feature vectors are pre-computed and reused during training, which reduces memory usage with greater efficiency.
3. We also designed a lightweight, residual connected MLP classifier containing fewer parameters than CNN fine-tuning, thereby reducing the computation and memory requirements. Because frozen CNNs provide transferable representations, the MLP can adapt these features efficiently for ransomware detection with high accuracy.

4. To rigorously assess the impact of the RL-based sample weighting, we benchmarked the framework against six baseline ML/DL classifiers (RF, XGBoost, SVM, LR, ANN, and CNN), both with and without RL. The RL-enhanced residual MLP achieved 99.1% accuracy and 99.74% AUC. We further profiled the training and inference times as well as memory consumption, demonstrating the computational efficiency of the framework. Furthermore, we employed t-SNE visualizations and Grad-CAM heatmaps to provide interpretable insights into feature separability and decision behavior.

The remainder of this paper is organized as follows. Section 2 reviews related works in detail, covering traditional ML, DL, TL, and RL paradigms. Section 3 elaborates on the methodology, detailing dataset, CNN-based feature extraction and fusion, RL-driven reweighting, and classifiers used. Section 4 outlines the experimental setup, evaluation metrics, dataset partitioning, and results. Section 5 offers a comprehensive discussion. Finally, Section 6 concludes the paper and suggests directions for future research.

**2. Related Works**

This section reviews prior research in three key areas that inform our proposed approach: (1) ransomware detection using deep learning, (2) transfer learning and multi-CNN feature fusion, and (3) reinforcement learning for adaptive sample reweighting in classification. This review integrates our research into the wider academic discourse, highlights unresolved issues, and identifies areas that require further investigation. **Table 1** compares the DL, TL, and RL approaches.

*2.1 Deep learning for ransomware detection*

The increasing complexity and evasion techniques of modern ransomware necessitate a transition from traditional signature-based detection methods to data-driven approaches. Recent studies have employed convolutional neural networks (CNNs), recurrent architectures, and hybrid deep models to classify ransomware by analyzing system logs, behavioral traces, and visual representations derived from ransomware binaries or execution patterns.

Manavi and Hamzeh introduced two DL approaches based on PE header byte images. The first approach [23] employed an LSTM model with two recurrent layers and a dense classifier, achieving 93.25% accuracy on 1,000 ransomware and 1,000 benign samples each. The second method [24] extracted headers from executables, converted them to grayscale images using zigzag patterns, and trained a CNN model with an accuracy of 93.33%. However, this method fails to detect new ransomware families. Moreira et al. [11] used static analysis to identify ransomware by transforming PE header files into color images in sequential vector format, classifying them using Xception CNN with 98.20% accuracy. They emphasized the importance of execution features for understanding advanced ransomware. Qin et al. [25] proposed an improved TextCNN model using dynamic feature such as API call sequences for ransomware detection, achieving an accuracy of 95.9%. Similarly, XRan [26] leveraged CNNs on dynamic features (API calls, DLLs, mutexes) integrated with XAI methods such as LIME and SHAP, achieving a true positive rate of 99.4%. However, this method is computationally intensive and time-consuming. Moreover, SwiftR [27] applied hierarchical neural networks across multiple platforms and demonstrated a robust cross-platform detection performance.

Collectively, these studies highlight the promise of deep learning for ransomware detection but also reveal persistent challenges in computational efficiency, dataset balance, and interpretability.

*2.2 Transfer learning for ransomware detection*

TL proves effective in contexts where domain-specific training data is limited. Pretrained CNNs such as VGG16, ResNet50, EfficientNet, and InceptionV3, originally trained on large-scale image datasets (e.g., ImageNet), can be adapted for malware image classification through techniques such as fine-tuning or feature extraction.

For example, Almomani et al. [28] introduced E2E-RDS, a vision-based ransomware detection system that transforms binaries into two-dimensional images, and evaluated 19 CNN models. A fine-tuned ResNet50 achieved 99.5% accuracy, outperforming the static ML baselines. Rezende et al. [29] used TL with ResNet-50 on grayscale byte plots, reporting 98.62% accuracy across 9,339 samples from 25 families. Although this study underscores the success of TL, it does not address the challenges of managing imbalanced datasets or new malware variants. Huang et al. [30] developed a hybrid model combining static and dynamic image representations, with VGG16 achieving 94.7% on hybrid data. However, this study did not evaluate the scalability or resource efficiency of the model. In our previous work [18], we developed a ransomware detection framework that converts sandbox features into grayscale and color images and applied fine-tuned CNNs, including ResNet50, EfficientNetB0, InceptionV3, Xception, VGG16, and VGG19. ResNet50 achieved 99.96% accuracy on 500 ransomware and 500 benign samples, respectively. However, fine-tuning requires substantial training time and memory, limiting its practicality in resource-constrained environments.

Subsequent studies have demonstrated the benefits of multi-backbone fusion. Shah et al. [14] introduced a dual-stage DWT with modified DenseNet-121 for feature extraction and machine learning classifiers, with XGBoost achieving 98.58% accuracy. The method optimized TL for grayscale inputs but lacked evaluation on unseen samples. Rustam et al. [15] proposed a Bi-model framework using VGG16 and ResNet50 as feature extractors, with traditional classifiers applied on fused features; their Bi-SVC achieved 100% accuracy on the Malimg dataset, but the work was limited to a single benchmark dataset, raising concerns of overfitting. Rezende et al. [16] used VGG16 and SVM to classify malware families with 92.97% accuracy, but underperformed newer ensembles. Kumar and Panda [13] proposed SDIF-CNN using VGG16, VGG19, ResNet50, and InceptionV3 as feature extractors, with an MLP classifier achieving 98.55% accuracy on Malimg and strong generalization on packed malware. Vasan et al. [12] developed IMCEC, an ensemble of pretrained CNNs with SVM and Softmax classifiers, reaching 99.5% accuracy on Malimg and showing resilience to obfuscation.

Collectively, these studies underscore TL's ability to deliver high accuracy and reduce manual feature engineering. However, most existing methods focus on single datasets and static visualizations without adaptive sample prioritization. Moreover, while accuracy is emphasized, efficiency (time, RAM), and model interpretability are less explored.

**Table 1** Comparative summary of related studies on ransomware detection. This table synthesizes prior approaches across DL, TL, and RL, highlighting feature extractors, feature type and analysis technique, classifiers, datasets, performance, strengths, and limitations.

| Category | Paper | Feature type and analysis technique | Feature Extractor(s)/ Classifier(s) | Dataset (R=Ransomware, B=Benign) | Accuracy (%) | Strengths | Limitations |
|---|---|---|---|---|---|---|---|
| Deep Learning (DL) | [23] | PE header bytes (static) | LSTM | R=1000, B=1000 | 93.25 | Simple DL pipelines, no feature engineering | Small dataset; no unseen families |
| | [24] | PE images (static) | CNN | R=1000, B=1000 | 93.33 | Simple DL pipelines, no feature engineering | Unable to detect unseen and evolving ransomware |
| | [11] | PE header color images (static) | Xception CNN | R=2023, B=2134 | 98.20 | Improved static image-based accuracy | Static features limit generalizability by missing ransomware behaviors. |
| | [25]. | API call sequences (dynamic logs) | Text CNN | R=1000, B=1000 | 95.9 | Effective on API behavior | Lacks interpretability |
| | [26] | API, DLL, mutex (dynamic) | CNN + XAI | R=668 B=668 | TPR 99.4 | High accuracy and explainability | Computationally intensive |
| | [27] | Hybrid features | HNN | R= 10.3K B=10 K | 98.75 | Robust across OS | Computational overhead |
| Transfer Learning (TL) | [28] | APK files images | 19 pre trained models | R=500 B=500 | 99.5 | Hybrid static and vision analysis | May struggle with modern evolving ransomware. |
| | [29] | Malware executables into images | ResNet-50 | Malimg dataset 9,339 malware samples | 98.62 | Robust TL on grayscale byteplots | No imbalance analysis |
| | [30] | Hybrid images (static and dynamic) | VGG16 | M=512 B=888 | 94.70 | Hybrid static and dynamic based detection | Limited scalability |
| | [18] | Image based dynamic features | ResNet50, EfficientNetB0, InceptionV3, | R=500 B=500 | 99.96 | State-of-the-art accuracy | High training time memory |
| | [14] | Grayscale images (static) | DenseNet-121 (feature extractor) XGBoost (classifier) | Malimg | 98.58 | Efficiency improved | No assessment was done on unseen or obscured samples. |
| | [15] | Malware binaries into images | VGG16 and ResNet50 (feature extractor), Bi-SVC (classifier) | Malimg | 100 | Bi-model architecture with stacked features enhances malware detection. | Raising concerns of overfitting. |
| | [16] | Byteplot grayscale images | VGG16 (feature extractor), SVM (classifier) | 10,136 malware samples | 92.97 | Removing feature engineering | Underperforming compared to newer ensembles. |
| | [13] | Malware binaries into images | VGG16, VGG19, ResNet50, and InceptionV3(feature extractor), Optimized MLP (classifier) | Malimg | 98.55 | Ensemble TL improves robustness and robust to packing | Computationally heavy |
| | [12] | Malware binaries into images | VGG16, ResNet50, and InceptionV3 (feature extractor), SVM (classifier) | Malimg and IoT-android mobile dataset | 98.82 | Robust to obfuscation and polymorphism | The authors overlooked the effects of image size and data imbalance on their method. Small and old dataset. |
| Reinforcement Learning (RL) | [19] | Static PE headers features | Double DQN | Dataset 1 had 62,485 PE files; Dataset 2 included 688 ransomware samples. | 97.9 | Early detection; avoids execution risk | Static headers vulnerable to packing and obfuscation |
| | [20] | Dynamic embeddings | RL + Bayesian Networks | Vulnerability and system logs | N/A | Adaptive resilience modelling | Not direct classification |
| | [21] | Android permissions and network traffic features | DRL (A2C, DQN, DDQN, PPO) | Android ransomware dataset | 89.78 | Improved accuracy, recall, F1 | Limited to static attributes |
| | [31] | device activity patterns | FL and RL | - | Above 98% | Enhancing adaptability | Scalability underexplored |

## 2.3 Reinforcement learning methods

RL has recently been explored for ransomware detection owing to its adaptability to evolving such threats. Unlike conventional ML, RL agents optimize policies via reward-driven interactions, making them suitable for dynamic threat environments.

Deng et al. [19] proposed an early ransomware detection framework that applied Double Deep Q-Learning (DDQN) to portable executable (PE) header features. This method achieved fast classification but was inherently limited by its reliance on static PE headers, which are prone to packing and are obfuscated. Kumar et al. [20] integrated RL with Bayesian networks and dynamic embeddings to strengthen resilience modelling across system configurations and vulnerabilities. Although effective for adaptive defense, this approach focuses on resilience prediction rather than direct ransomware classification. Jeremiah et al. [21] used Deep Reinforcement Learning (A2C, DQN, DDQN, PPO) on Android permissions and network traffic. Their A2C and DDQN models surpassed traditional ML baselines in accuracy and speed, though they only considered static features rather than behavioral representations. CyberForce uses federated reinforcement learning (FRL) for IoT malware mitigation, achieving 98% accuracy and reducing training time by 67% versus centralized methods [31].

While RL shows promise in adaptability and performance, studies are limited to static PE features and Android contexts, with minimal focus on image-based ransomware or accuracy-efficiency optimization.

This comprehensive synthesis of the existing literature reveals that while each paradigm ML, DL, TL, and RL offers unique advantages, none individually addresses the full set of requirements for modern ransomware detection. Traditional ML methods fail to generalize beyond handcrafted features, whereas DL methods achieve high accuracy but require large datasets and computational resources. TL addresses these issues by leveraging pretrained models; however, most TL-based approaches remain non-adaptive and treat all samples equally, leading to overfitting on trivial patterns while neglecting hard-to-detect behaviors. Although TL is increasingly used in malware analysis, the integration of RL to adaptively reweight training samples has not been explored in the ransomware domain. To the best of our knowledge according to the literature, no prior study has integrated TL with RL in an image-based ransomware detection pipeline. This motivated the development of TL-RL-FusionNet, which bridges these gaps by (a) transforming dynamic behavioral features into images, (b) extracting complementary embeddings through frozen EfficientNetB0 and InceptionV3 backbones, and (c) training a lightweight residual MLP under a Q-learning-based sample-weighting agent.

## 3. Methodology

This section outlines the comprehensive methodology employed in the study, encompassing dataset preparation, preprocessing, and the design of the proposed TL-RL-FusionNet framework. **Fig. 1** illustrates the overall architecture and operational workflow of the proposed end-to-end framework for detecting ransomware. The methodology is systematically organized into six components: (i) Dataset preparation, (ii) Data preprocessing, (iii) Feature Conversion, (iv) Dual-CNN feature extraction and fusion, (v) Reinforcement learning-guided training, and (vi) Description of classifiers utilized in this study.

### 3.1 Dataset

This study was evaluated using a custom dataset created in our previous study [18]. It contains 1,000 binary executables, comprising 500 ransomware and 500 benign programs collected from various online repositories. All ransomware samples were collected from MalwareBazaar [32] and VirusShare [33] and selected from 25 distinct ransomware families, as classified by threat intelligence. Benign samples were obtained from trusted repositories such as SnapFiles [34], PortableApps.com [35], and GitHub [36], which ensured low label noise and high dataset integrity. Certain criteria were applied to select both the ransomware families and individual ransomware samples within each family to ensure the dataset's relevance, diversity, representativeness, and operational significance.

Two criteria were applied for the selection of the ransomware family:

1. Its high prevalence and significant global impact on organizations have been documented in multi-source threat intelligence reports from 2019 [37], 2020 [38], 2021 [39], 2022 [40], 2023 [41], and 2024 [42].
2. Verification across at least two independent reports from reputable cybersecurity vendors.

This selection method aimed to identify prominent ransomware families, such as LockBit, MedusaLocker, BlackCat, Phobos, and Conti, to ensure that the dataset included representative coverage of advanced and adaptive threats. Prioritizing these high-impact variants enhances the understanding of critical ransomware behaviors and increases the relevance of the dataset for detection models.

Individual ransomware samples within each family were selected following the procedure described in [11], with three criteria applied based on the VirusTotal vendor engine detection. A ransomware sample was included if it met all the following criteria:

1. At least 45 antivirus engines on VirusTotal classified the files as malicious.
2. At least 15 antivirus engines specifically labelled the files as ransomware.
3. The majority of engines or at least ten, identified the file as belonging to the same ransomware family.

Each sample was executed in an isolated Cuckoo Sandbox environment [43], and reports were generated in JSON format, which documented both the ransomware attack lifecycle and the normal footprint of benign software, forming the basis for feature extraction. A custom Python-based pipeline was then used to parse these reports into structured, fixed-length feature vectors. From each sample, 100 highly discriminative behavioral indicators were selected, spanning file-system changes, registry modifications, network activity, process and memory operations, API call patterns, and anti-analysis artifacts. The selected features were cross-referenced with MITRE ATT&CK tactics and validated against threat reports from Sophos, CrowdStrike, and Trend Micro (2019–2024). This curated set ensures technical depth, practical relevance, and improved model detection performance across ransomware families.

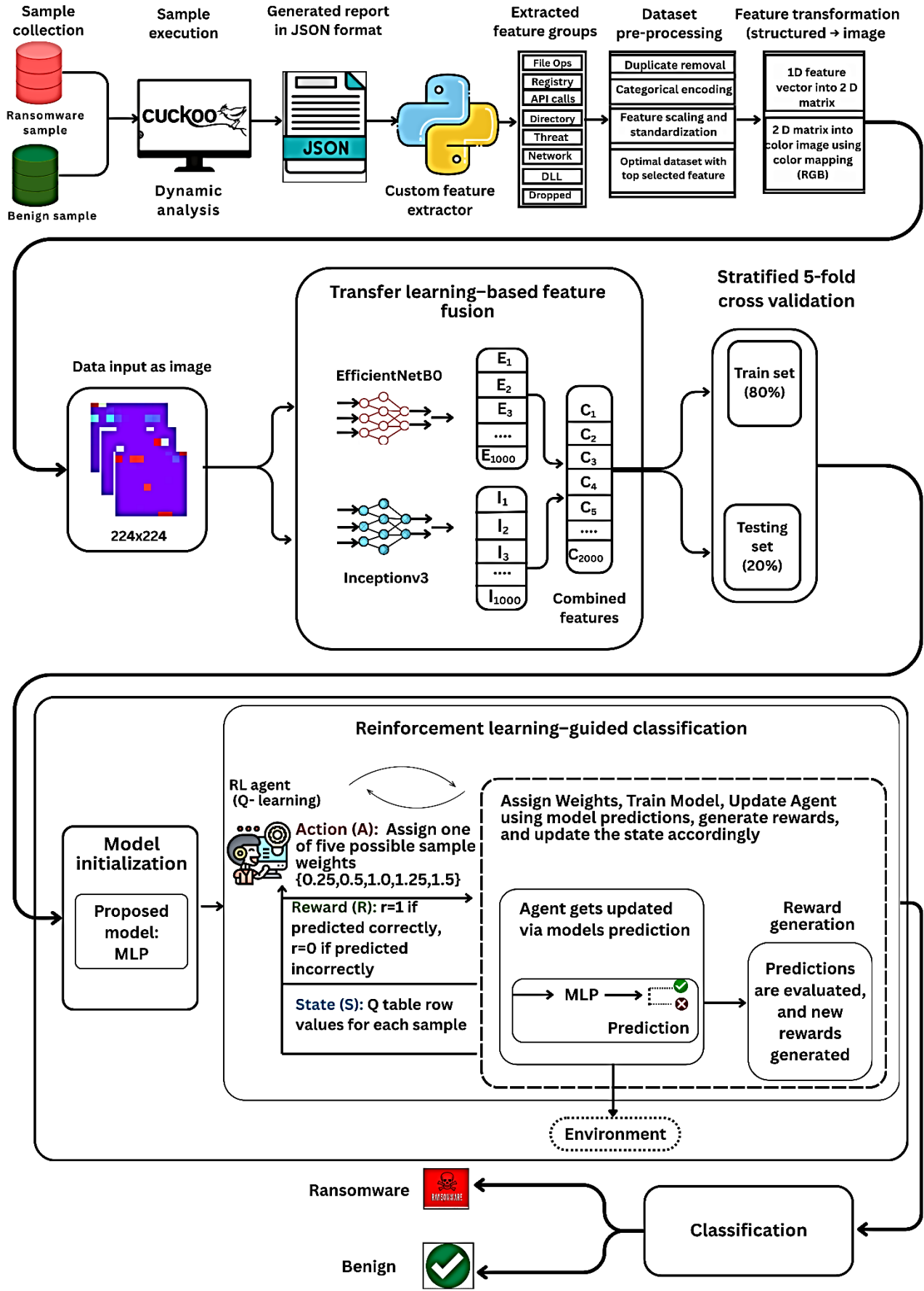

**Fig. 1.** End-to-end TL-RL-FusionNet workflow for ransomware detection.

**Table 2** presents the detailed distribution of the ransomware families included in this study and the corresponding number of samples.

**Table 2** Ransomware families included in this study, with sample counts. Families were selected based on global prevalence, documented impact, and multi-vendor confirmation of classification.

| Family name | No. of samples |
|---|---|
| Ryuk | 15 |
| GandCrab | 28 |
| Cerber | 9 |
| Maze | 8 |
| Makop | 12 |
| RagnarLocker (Ragnarok) | 23 |
| NetWalker | 24 |
| Mespinoza (Pysa) | 18 |
| Clop | 10 |
| REvil (Sodinokibi) | 26 |
| LockBit | 30 |
| Phobos | 24 |
| WastedLocker | 25 |
| DarkSide | 24 |
| Avaddon | 16 |
| MountLocker | 15 |
| Thanos | 28 |
| Conti | 18 |
| Nefilim | 11 |
| BlackCat | 30 |
| Babuk | 27 |
| BlackMatter | 28 |
| MedusaLocker | 28 |
| Mallox | 15 |
| BlueSky | 8 |

### 3.2 Data preprocessing

Data preprocessing is a critical step in ensuring the quality, consistency, and compatibility of machine learning training and evaluation. In this study, three key preprocessing steps were applied: removal of duplicate records, encoding of categorical attributes into numeric form, and feature scaling and standardization. The result was a clean, fully numeric dataset that was prepared for subsequent analysis and model development.

1. *Duplicate removal*: Duplicate records from repeated executions or redundant reports were removed to prevent bias and ensure reliable generalization of the model.
2. *Categorical encoding:* Non-numeric fields were transformed into numerical values to ensure compatibility with the machine learning models.
   i) Boolean indicators (e.g., privilege escalation, shadow copy deletion, evasion) were encoded as binary (1/0); (ii) string attributes (e.g., IP origin, country) were mapped to numeric values based on length or frequency; and (iii) list-based features (e.g., dropped files, DNS requests, VirusTotal positives) were represented by element counts.
3. *Feature scaling and standardization*: Normalization was applied to ensure that all features operated on a uniform scale, thereby improving the classifier convergence and stability.
4. *Updated optimal dataset with top selected features:* the optimized dataset comprises the top 100 high-discriminative features selected through feature selection and validated via MITRE ATT&CK and threat intelligence mapping, effectively reducing dimensionality and computational cost while preserving maximal detection accuracy.

This preprocessing pipeline ensured a fully numeric, standardized dataset ready for structured-to-image transformation and classification in TL-RL-FusionNet.

### 3.3 Conversion of structured features into image

To enable CNN-based transfer learning, the one-dimensional feature vectors were reshaped into 224×224 matrices and encoded as RGB images using a colormap. This transformation preserved the inter-feature relationships while providing a visual domain for pretrained CNN backbones to extract deep representations.

### 3.4 Transfer learning for feature fusion from image data

Two pretrained CNNs, EfficientNetB0 and InceptionV3, were used as frozen feature extractors. Both models were pretrained on ImageNet, and their convolutional layers were frozen to preserve powerful, generalized image representation capabilities. Their selection was driven by two key factors.

1. *Complementary strengths:* EfficientNetB0 applies compound scaling to balance the resolution, width, and depth efficiently, making it highly suitable for extracting fine-grained ransomware behaviors with a low parameter count. InceptionV3 incorporates inception modules that capture semantic patterns at multiple scales, complementing the local detail focus of EfficientNet. For each input image, both CNNs generated global average pooled feature embeddings. To enhance the richness and discriminability of the representations, the pooled outputs from EfficientNetB0 and InceptionV3 were concatenated into a single fused feature vector. This fusion leverages the complementary learning characteristics of both the CNN architectures. The resulting combined vector captures a broader spectrum of spatial, texture, and frequency features, enabling a stronger classification performance downstream.

2. *Proven success in ransomware detection:* Image-based CNN pipelines consistently achieve state-of-the-art accuracy in ransomware detection. In our previous study [18], ResNet50 achieved the highest accuracy (99.96%), followed by EfficientNetB0 (99.78%) and InceptionV3 (97.84%). Although ResNet50 performed the best, its training cost was significantly higher. EfficientNetB0 and InceptionV3 offered the optimal trade-off between accuracy and efficiency, making them the most practical choices for dual frozen feature extraction in the TL-RL-FusionNet. Their selection is further supported by the proven success in the broader ransomware detection literature, where both models have demonstrated strong generalization and effectiveness across diverse datasets.

### 3.5 Reinforcement learning-based sample weighting

RL is a machine learning paradigm in which an agent learns to make sequential decisions through trial-and-error interactions with an environment, aiming to maximize the cumulative reward. Unlike supervised learning, RL does not rely solely on labelled datasets; instead, it learns optimal strategies through feedback from the environment [19].

In ransomware detection, RL can guide a model to emphasize the most informative samples during training. Rather than treating all data points equally, the RL agent assigns varying weights to the training samples, enabling the classifier to focus on examples that are more influential in improving the decision boundaries. This targeted weighting helps the model learn patterns that are more representative of ransomware and benign behaviors, even when these patterns are subtle or imbalanced in the dataset.

The general interaction loop between the RL components is illustrated in **Fig. 2**, which highlights the flow of states, actions, and rewards [44]. At each time step t, the agent observes the current state St, selects an action At, receives a reward Rt from the environment based on the classification performance, and transitions to the next state St+1. Over time, the agent optimizes its decision policy by balancing exploration of new weighting strategies with exploitation of those that have high-reward actions.

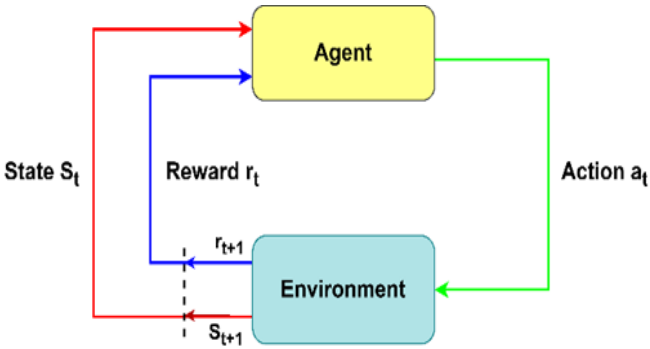

**Fig. 2.** Basic Reinforcement Learning components and information flow between Agent and Environment, showing the sequence of state observation, action selection, reward feedback, and state transition.

*3.5.1 RL component mapping in our study*

In our ransomware detection framework, the environment is the model training dataset in each cross-validation fold. The RL integration proceeds as follows.

- *Agent:* The agent is a Q-learning–based sample weighting module
- *Environment:* Environment is the ransomware classification model and its performance on the training split within each cross-validation fold.
- *State (St):* Q-table row values representing the learned weight preferences for each training sample.
- *Action (At):* Selection of one of five discrete sample weight values {0.25, 0.5, 1.0, 1.25, 1.5}
- *Reward (Rt):* Assigned as +1 if the sample is correctly classified by the current model during training or 0 if misclassified.
- *Policy update:* Q-values are updated using the standard Q-learning update rule, based solely on test predictions.

In the context of our ransomware detection framework, RL offers the following two key advantages:
1. *Adaptive focus on hard samples*: The agent can assign greater importance to ransomware samples that the classifier struggles with, reducing the misclassification risk.
2. *Dynamic adjustment over epochs*: Sample weights are updated iteratively based on the ongoing model performance, allowing the system to adapt to evolving decision boundaries.

*3.5.2 Integration into model training pipeline*

The RL mechanism described here is applied within the cross-validation framework defined in Section 4.3. For every fold in the stratified 5-fold CV, the RL agent was re-initialized to ensure that no learned weighting policy from previous folds influenced the current evaluation. All Q-value updates and reward assignments are computed exclusively from the training split of that fold, completely excluding the test set predictions. This integration guarantees that the adaptive weighting process is learned under the same conditions as the model training, without any data leakage.

**Operational flow per fold in CV**

*1) Agent initialization:* At the start of each fold, the reinforcement learning (RL) agent was initialized with a fresh Q-table of shape $(N,5)$, where $N$ is the number of training samples. All the Q-values were set to zero. The discrete action set of possible sample weight multipliers is:

$$A = \{0.25, 0.50, 1.00, 1.25, 1.50\} \quad (1)$$

*2) State representation:* Each training sample $i$ is treated as a state $S_i$, represented by the $i^{th}$ row of the Q-table as follows:

$$Q(S_i, \cdot) \in \mathbb{R}^5 \quad (2)$$

*3) Action selection (weight assignment):* For each training sample $i$, the agent selects an action $a_i$ using an $\epsilon$-greedy policy as follows:

$$a_i = \begin{cases} \text{uniform}(A) & \text{with probability } \varepsilon \text{ (exploration)} \\ \arg\max_{a \in A} Q(S_i, a) & \text{with probability } 1 - \varepsilon \text{ (exploitation)} \end{cases} \quad (3)$$

The selected action directly sets the sample training weight as follows:

$$w_i = a_i \in A \quad (4)$$

After weights are assigned for the fold, the agent decays its exploration rate as follows:

$$\varepsilon \leftarrow \varepsilon \cdot \epsilon_{decay} \quad (5)$$

*4) Reward calculation (as implemented):* After training the classifier on the weighted training data, predictions $\hat{y}_i$ were generated for the validation split. The reward for each validation sample $i$ is as follows:

$$r_i = \begin{cases} 1 & if\ \hat{y}_i = y_i, \\ 0 & \text{otherwise.} \end{cases} \quad (6)$$

This design uses validation predictions to guide the Q-table updates, ensuring that the weighting adjustments are influenced by the model performance on unseen data within the fold, thus encouraging generalizable weighting strategies.

*5) Q-value update (same-state bootstrapping):* The Q-value for the executed action is updated as follows:

$$Q(S_i, a_i) \leftarrow Q(S_i, a_i) + \alpha[r_i + \gamma \max_{a \in A} Q(S_i, a) - Q(S_i, a_i)] \tag{7}$$

where $\alpha$ is the learning rate and $\gamma$ is the discount factor. Bootstrapping is performed using the same state $S_i$, which is consistent with the implementation.

*6) Weighted training objective:* In this phase, the model parameters were optimized using a weighted loss function, in which each training sample was assigned an importance weight by a reinforcement learning (RL) agent. These RL-assigned weights enable the model to focus on samples that contribute more significantly to learning, thereby improving generalization and model efficiency. Training minimizes the weighted mean loss used by Keras as follows:

$$L_{RL} = \frac{\sum_{i=1}^{N} w_i l(y_i, \hat{y}_i)}{\sum_{i=1}^{N} w_i}, \tag{8}$$

where $\ell(\cdot)$ is the base loss (categorical cross-entropy with label smoothing), $w_i$ is the RL-assigned weight, and $\hat{y}_i = f_\theta(x_i)$ is the model prediction for input $x_i$.

The optimal model parameters were obtained as follows:

$$\theta^* = \mathrm{argmin}_\theta L_{RL} \tag{9}$$

where $\theta* =$ is the optimal set of model parameters after training that minimizes the total weighted loss over all N training samples, and argmin$\theta$ = "Argument of the minimum" it means we search for the $\theta$ that makes the expression inside have the smallest value.

The complete process is outlined in Algorithm 2, which describes the update mechanism for the RL agent in each fold.

---

**Algorithm 2: RL-Based Dynamic Sample Weighting Agent**

**Input:**
  Number of samples $N$
  Action space $\mathcal{A} = \{0.25, 0.5, 1.0, 1.25, 1.5\}$
  RL parameters: $\alpha$ (learning rate), $\gamma$ (discount), $\varepsilon$ (exploration), $\varepsilon_{decay}$
  Ground truth labels $y$ and model predictions $\hat{y}$

**Output:**
  Sample weights $w \in \mathbb{R}^N$ (for the training set $T$)
  Updated Q-Table $Q \in \mathbb{R}^N \times |\mathcal{A}|$

**1. Initialization:**
  $Q \leftarrow$ zeros $(N, |\mathcal{A}|)$; $a_i \leftarrow 0 \ \forall_i \in \{1, \ldots, N\}$

**2. Assign Weights (Training Phase):**
  For each sample $i \in T$:
    Choose action index $a$ by $\varepsilon$-greedy over $Q(i, \cdot)$
    Set $a_i \leftarrow a$
    Assign weight $w_i \leftarrow \mathcal{A}[a]$
  Update exploration: $\varepsilon \leftarrow \varepsilon \times \varepsilon_{decay}$

**3. Update Q-table (Validation Phase):**
  For each sample $i \in V$:
    Reward $r \leftarrow \begin{cases} 1, & \hat{y}_i = y_i \\ 0, & \text{othersise} \end{cases}$
  Update:
  $Q[i, a_i] \leftarrow Q[i, a_i] + \alpha \ (r + \gamma \cdot \max_a Q[i, a] - Q[i, a_i])$

**Notes:**
  $a_i$ stores the last action index associated with sample $i$; if no action was assigned previously, $a_i = 0$
  (Default weight: 0.25).
  Training minimizes the Keras-weighted mean loss: $L_{RL} = \frac{\sum_{i=1}^{N} w_i l(y_i, \hat{y}_i)}{\sum_{i=1}^{N} w_i}$

---

*3.6 Classifiers used in this study*

Although the RL-weighted MLP represents the core of our proposed approach, we benchmarked its performance against six other classifiers, both with and without RL-based weighting. These models were selected to represent a diverse range of learning paradigms, from traditional to ensemble and deep learning models.

*3.6.1 Residual MLP model development (proposed)*

The Residual MLP is a specialized feed-forward neural network designed to process fixed-length feature vectors extracted from pretrained CNNs. Unlike conventional multilayer Perceptrons, the Residual MLP integrates residual skip connections within its hidden layers to improve gradient propagation, alleviate the vanishing gradient problem, and enable stable training of deeper architectures. This design ensures that the model maintains a high representational capacity while avoiding the degradation in performance often observed when additional layers are added to a standard MLP. An architectural diagram of the residual MLP for ransomware classification is shown in **Fig. 3**.

The network begins with a fully connected (dense) layer comprising 1,024 hidden units, followed by batch normalization to stabilize the learning dynamics and dropout regularization to mitigate overfitting. The core of the architecture consists of two Residual Blocks, each implementing a bottleneck structure. In each block, the input passes through a dense layer with 256 units and a Rectified Linear Unit (ReLU) activation, followed by batch normalization and another dense layer that restores the dimensionality to 1,024 units. A skip connection adds the original input of the block to its transformed output, after which a rectified linear unit (ReLU) activation is applied. A dropout was introduced after each block to further reduce overfitting. The final output layer applies a softmax activation for binary classification (ransomware vs. benign). The complete architectural values are summarized in **Table 3**, and the training configuration and hyperparameters are provided in **Table 4.**

The classification process of the proposed lightweight MLP is detailed in the following section, including its linear transformations, nonlinear activations, residual connections, normalization, and regularization strategies.

*Input processing and dense layer transformation:* The classification process begins with a concatenated feature vector extracted from multiple CNN backbones (EfficientNetB0, InceptionV3) that undergoes a linear transformation. Dense layers perform linear transformations of the input vector x using a weight matrix W and bias vector b, followed by an activation function, ReLU, which is used in the hidden layers. Hence the linear output (z) can be determined as-

$$z = W_x + b \tag{10}$$

ReLU(z) is defined as

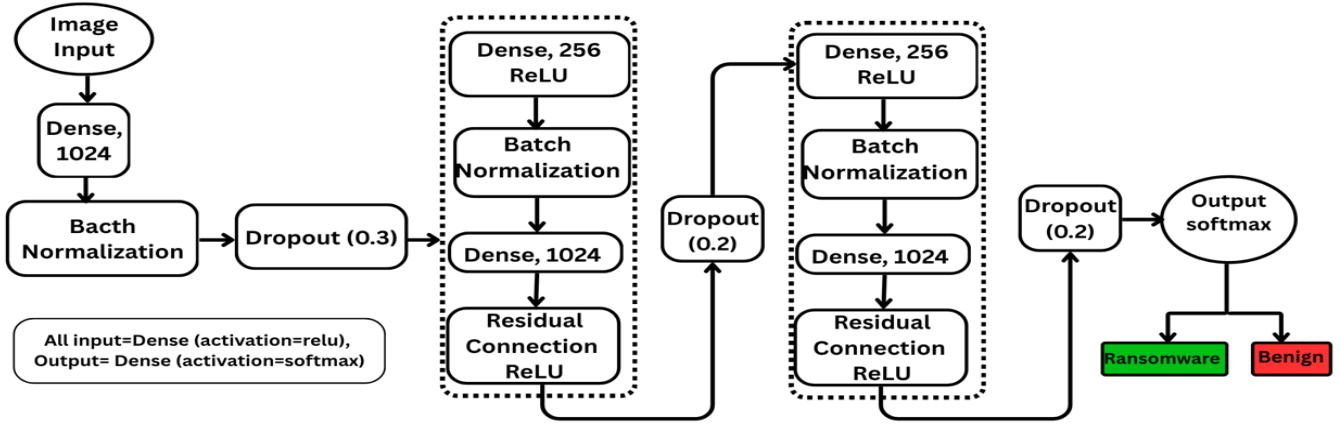

**Fig. 3.** Architectural diagram of residual MLP

**Table 3** Residual MLP architectural values

| Layer | Output Shape | Parameters |
|---|---|---|
| Input | (None, 3328) | 0 |
| Dense | (None, 1024) | 3,408,896 |
| BatchNormalization | (None, 1024) | 4,096 |
| Dropout (0.3) | (None, 1024) | 0 |
| Residual Block ×2 | (None, 1024) | 1,053,184 |
|    Dense (256) | (None, 256) | 262,400 |
|    BatchNormalization | (None, 256) | 1,024 |
|    Dense (1024) | (None, 1024) | 263,168 |
|    Add + ReLU | (None, 1024) | 0 |
|    Dropout (0.2) | (None, 1024) | 0 |
| Output Dense (Softmax, 2) | (None, 2) | 2,050 |
| Total Parameters | | 4,468,226 |

**Table 4** Training and hyperparameters

| Parameter | Value |
|---|---|
| Activations | ReLU (hidden), Softmax (output) |
| Dropout Rates | 0.3 (pre-block), 0.2 (inside block) |
| Normalization | BatchNorm after each Dense |
| Optimizer | Adam |
| Learning Rate | Cosine annealing, init 1e-3 |
| Loss | Categorical cross-entropy with label smoothing 0.1 |
| Batch Size | 32 |
| Epochs | 25 |
| Cross-validation | 5-fold stratified |
| Sample Weighting | RL ε-greedy; Actions = {0.25, 0.5, 1.0, 1.25, 1.5}; Reward = 1 if correct, else 0 |

$$ReLU(z) = \begin{cases} z, & \text{if } z \text{ is positive } (z > 0), \text{it stays the same} \\ 0, & \text{if } z \text{ is negative } (z \leq 0), \text{it becomes } 0 \end{cases} \quad (11)$$

The hidden layers apply the ReLU activation to the linear output as follows:

$$ReLU(z) = \max(0, z) \quad (12)$$

Combining the two steps, the dense layer transformation for the hidden layers can be expressed as

$$Dense(x) = ReLU(W_x + b) = \max(0, W_x + b) \quad (13)$$

This formulation ensures nonlinear feature mapping while maintaining computational efficiency, as ReLU introduces sparsity by zeroing negative activations.

*Residual Connections:* Following the Dense transformations, each residual block implements identity shortcuts that add the input vector x to the block's transformed output $F(x, W)$ before applying activation:

$$y = ReLU(F(x, W) + x) \quad (14)$$

Here, $F(x, W)$ denotes the transformation consisting of the Dense–BatchNorm–Dense sequence within a block. This additive formulation allows the network to learn modifications to the identity mapping rather than the complete transformation, facilitating a faster convergence and improved generalization.

*Regularization strategy:* Regularization was applied in two forms.
Dropout: A dropout rate of 0.3 was applied immediately after the initial dense layer, and a rate of 0.2 was applied after each residual block. This stochastically removes a fraction of the units during training, forcing the network to learn redundant representations and reducing overfitting.
Batch normalization: Batch normalization is performed after every dense layer to normalize intermediate activations, reduce internal covariate shift, and enable the use of higher learning rates without divergence.

*Softmax output layer:* For the output layer, a Softmax activation is applied to produce the probabilities for each class. For class $i$, the probability is given by

$$\hat{y}_i = \frac{e^{z_i}}{\sum_{j=1}^{C} e^{z_i}} \quad (15)$$

By substituting $z = Wx + b$, we

$$\hat{y}_i = \frac{e^{(Wx+b)_i}}{\sum_{j=1}^{C} e^{(Wx+b)_i}} \quad (16)$$

where C is the number of classes and z is the input to the softmax function. For binary classification $(C=2)$, softmax reduces to the logistic sigmoid for one output neuron, providing interpretable class probabilities.

*RL-guided classification:* During training, the RL agent dynamically adjusted the sample weights to prioritize harder or more informative examples. The agent employs an ε-greedy policy over discrete weight multipliers {0.25, 0.5, 1.0, 1.25, 1.5}, updating the Q-values based on whether the predictions are correct (reward = 1) or incorrect (reward = 0) for the training split. This ensures that the model increasingly focuses on samples that yield the highest performance gains during training.

Training and optimization: The network were trained using the Adam optimizer with a cosine annealing learning rate schedule, starting from an initial learning rate of $1\times10^{-3}$. The categorical cross-entropy loss function was employed with a label smoothing factor of 0.1 to encourage the model to be less overconfident and improve its ability to generalize. The model was trained for 25 epochs with a batch size of 32 under a 5-fold stratified cross-validation protocol to ensure robust performance estimation.

Algorithm 3 summarizes only the forward classification pipeline of the residual MLP from the feature vector input to the probability estimation.

**Algorithm 3: Lightweight Residual MLP Classifier**

**Input:** Feature vector $x \in \mathbb{R}^d$
**Output:** Softmax probability vector $\hat{y} \in \mathbb{R}^c$

1. **Input Layer:**
   $h_0 \leftarrow$ Dropout (0.3) [ BatchNorm (ReLU (Dense$_{1024}$(x)))]
2. **Residual Block × 2:**
   For each block:
     res ← h
     h ← Dense$_{256}$ → ReLU → BatchNorm → Dense$_{1024}$
     h ← Dropout (0.2) (ReLU (h + res))
3. **Output Layer:**
   $\hat{y} \leftarrow$ Softmax (Dense_C(h))
4. **Training:**
   Optimizer: Adam with cosine annealing ($\eta_0 = 1\times10^{-3}$)
   Loss: categorical cross-entropy with label smoothing ($\alpha = 0.1$)
   Batch size = 32, Epochs = 25, 5-fold cross-validation

*3.6.2 Baseline models*

To evaluate the effectiveness of the proposed RL-guided MLP model, we selected a set of diverse and widely used baseline models spanning traditional machine learning and deep learning paradigms. These include RF, XGBoost, SVM, LR, ANN and CNN. Each model in the benchmark set was carefully selected based on its relevance in the ransomware classification literature, compatibility with the transformed feature space (CNN-extracted high-level embeddings), and ability to provide complementary perspectives on learning behavior under RL-based sample weighting. Because the dataset consists of ransomware and benign behavior represented as color images, feature extraction was performed using EfficientNetB0 and InceptionV3, whose embeddings were concatenated to form a high-dimensional and discriminative feature vector. This selection ensured diverse methodological paradigms that aligned well with the characteristics of the dataset and proposed method.

Tree-based ensembles (RF, XGBoost) are well-suited because they handle high-dimensional, structured embeddings efficiently and provide robustness against feature noise, enabling the capture of nonlinear interactions among CNN-extracted features. Margin-based and linear learners (SVM, LR) offer complementary perspectives by leveraging the inherent separability of the CNN embeddings. While SVM exploits high-dimensional margins with kernel functions, LR provides a linear baseline to test whether the features are linearly separable. Neural networks (ANN and CNN) extend the evaluation to deep learning paradigms. ANN leverages dense layers to capture higher-order dependencies within concatenated embeddings, whereas CNN assesses whether spatial locality from image representations is preserved and beneficial after feature fusion.

This diversity ensures that the proposed RL-guided MLP model is benchmarked against classical and interpretable models and advanced deep learners, establishing both fairness and contextual relevance.

The working processes of these classifiers, including their training and prediction mechanisms, are described in the following sections.

*Random Forest (RF):* Random Forest is an ensemble learning method that constructs multiple decision trees using bootstrapped subsets of the training data and random subsets of features at each split. The final prediction is obtained by aggregating the outputs of all trees through majority voting, thereby reducing the variance and enhancing the robustness against overfitting [45].

The prediction of a Random Forest with M trees can be expressed as

$$\hat{y} = mode\{h_m(x) \mid m = 1, 2, \dots, M\} \qquad (17)$$

where $h_m(x)$ is the prediction of the $m^{th}$ decision tree.

By aggregating multiple weak learners, Random Forest achieves stable performance on high-dimensional CNN-derived features, making it effective for classifying ransomware, as it leverages the robustness of ensemble methods to improve detection accuracy.

*Support Vector Machine (SVM):* The SVM classifier is a supervised ML model that divides an n-dimensional space-based representation of the data into two classes using a selected boundary known as a hyperplane. This hyperplane maximizes the margin between the two regions or classes (in our experiment, ransomware, or benign software). The maximal margin is defined as the largest distance between examples of both classes computed from the distance between their closest instances (called supporting vectors) [5]. It is designed to maximize class separability while minimizing the classification error using slack variables.

Formally, the optimal hyperplane is represented by a vector $w$ and a scalar $m$ such that the inner products of $w$ with vectors $\phi(x)$ from the two classes are divided by an interval between $-1$ and $+1$ subject to b. Prediction is given by:

$$\hat{y} = sign(w^T \phi(x) + b) \qquad (18)$$

where $w^T$ denotes the transpose of the weight vector $w$.

The ability of SVM to handle high-dimensional fused features makes it a strong baseline for distinguishing ransomware from benign classes, particularly when dual-CNN embeddings increase feature separability.

*Extreme Gradient Boosting (XGBoost):* XGBoost is an advanced ensemble learning technique that enhances the traditional gradient boosting algorithm by integrating strong regularization terms to prevent overfitting in the model. Its scalability and efficiency make it particularly effective for high-dimensional features. XGBoost builds models sequentially, where each new learner attempts to reduce the residual errors of the previous ensemble [46].
The learning objective combines a data fitting loss with a regularization penalty:

$$L(\theta) = \sum_{i=1}^{n} l(y_i, \hat{y}_i) + \sum_{k=1}^{K} \Omega(f_k) \quad (19)$$

Here, $L(\theta)$ represents the overall regularized objective, $l(y_i, \hat{y}_i)$ is the loss between the true and predicted labels, $K$ is the total number of boosting trees, and $\Omega(f_k)$ is the regularization component that penalizes model complexity. The regularization function is typically defined as

$$\Omega(f_k) = \gamma^T + \frac{1}{2}\lambda ||w||^2 \quad (20)$$

where $T$ is the number of leaves in the decision tree, $\gamma$ is the regularization coefficient that penalizes the number of leaves, and $\lambda$ is the regularization parameter applied to the leaf weights $w$.

The final prediction was obtained by aggregating the outputs of all boosted trees:
$$\hat{y} = \sum_{m=0}^{M} \eta h_m(x) \quad (21)$$

where $M$ denotes the total number of trees, $h_m(x)$ is the prediction from the $m^{th}$ tree, and $\eta$ is the learning rate that controls the contribution of each tree.
XGBoost constructs a robust model that effectively minimizes the loss function and mitigates overfitting, thereby enhancing its efficacy for the classification task of Ransomware Detection.

*Logistic Regression (LR):* LR is a linear classifier that estimates the probability of a sample belonging to a particular class using the logistic (sigmoid) activation function. It is mainly used in binary classification tasks, using a logistic function (0-1) to estimate probabilities [47]. It is typically trained using maximum likelihood estimation (MLE) by minimizing the cross-entropy (log-loss) function.
For the binary classification of ransomware ($y$=1) versus benign ($y$=0) based on the fused feature embeddings, the probability function is defined as

$$\hat{P}(y = 1|x) = \sigma(w^T x + b) = \frac{1}{1+e^{-(w^T x+b)}} \quad (22)$$

where $P(y=1|x)$ is the predicted probability that $x$ belongs to the ransomware class, x is the feature vector, $w$ is the weight vector learned during training, $b$ is the bias term, $w^T x$ is the linear combination of features, and $\sigma(\cdot)$ is the sigmoid activation function.
Logistic Regression provides a simple yet effective baseline classifier to evaluate whether ransomware and benign samples are linearly separable in the feature space. Although more complex models (e.g., RL-guided residual MLP or XGBoost) can capture nonlinear interactions, LR offers interpretability and serves as a benchmark against which more advanced classifiers can be compared.

*Artificial Neural Network (ANN):* Artificial Neural Networks consist of multiple fully connected layers that can learn complex nonlinear mappings from inputs to outputs. Each hidden layer transforms its input through a weighted linear combination, followed by a nonlinear activation [47].
For the $l^{th}$ hidden layer,

$$h^l = \sigma(W^{(l)} h^{(l-1)} + b^{(l)}) \quad (24)$$

where $h^l$ is output (activation vector) of the $l^{th}$ hidden layer, $h^{(l-1)}$ is input vector from the $(l-1)^{th}$ layer $W^{(l)}$ is the weight matrix mapping inputs to hidden layer $l$, $b^{(l)}$ is the bias vector of hidden layer $l$, $\sigma(\cdot)$ is the nonlinear activation function (e.g., ReLU)
The final output layer applies the Softmax activation to produce class probabilities for binary classification (ransomware vs. benign):

$$\hat{y} = \text{softmax}(W^L h^{(L-1)} + b^L) \quad (25)$$

Where, $\hat{y}$ is predicted probability distribution over the two classes, $W^L$ is weight matrix of the final output layer, $b^{(L)}$ is bias vector of the final layer, $h^{(L-1)}$ is the output of the last hidden layer before classification.

The network parameters were optimized via backpropagation to minimize the categorical cross-entropy. ANN captures nonlinear interactions within the fused features, making it suitable for ransomware detection tasks.

*Convolutional Neural Network (CNN):* Convolutional Neural Networks are designed to capture local spatial patterns in images through convolutional layers, followed by pooling, nonlinear activations, and fully connected layers [9][48]. The convolutional operation is defined as:
$$h_{i,j}^{(k)} = \sigma(\sum_{p,q} X_{i+p, j+q} \cdot K_{p,q}^{(k)} + b^{(k)}) \quad (26)$$

Where $X$ is the input image, $K^{(k)}$ is the convolutional kernel, and $h^{(k)}$ is the resulting feature map. The final classification was performed using a softmax layer:

$$\hat{y} = \text{softmax}(Wh + b) \quad (27)$$

Where, $\hat{y}$ = predicted probability distribution across the two classes (ransomware, benign)
$W$ = weight matrix of the final fully connected layer
$h$ = flattened feature vector after convolution + pooling operations
$b$ = bias vector for the classification layer

## 4. Experiment and results

This section outlines the lab setup, dataset splitting and evaluation strategy, evaluation metrics, and result analysis used to validate the TL-RL-FusionNet framework. Our experiments were designed to assess the efficiency and robustness of the RL-driven sample weighting when combined with dual-backbone CNN feature extraction and MLP classification.

## 4.1 Experimental setup

The experimental setup of this study integrated advanced tools and techniques to detect ransomware using dynamic analysis methods. The analysis environment employed a Cuckoo Sandbox [43] was used to securely execute and monitor the ransomware samples, generating detailed behavioral reports for subsequent processing and feature extraction. The implementation was performed using Python 3.10.12, utilizing a comprehensive suite of libraries for deep learning, classical machine learning, data processing, and visualization. TensorFlow and Keras were used to design, train, and evaluate the deep learning models, leveraging pretrained architectures such as EfficientNetB0 and InceptionV3 for transfer learning. Scikit-learn facilitated classical machine learning algorithms (RF, SVM, KNN) and provided essential utilities for cross-validation, preprocessing, and performance evaluation metrics (accuracy, precision, recall, F1-score, ROC, and AUC). XGBoost was used for gradient-boosting-based classification. NumPy and Pandas supported numerical operations and structured data handling, whereas rarfile enabled dataset extraction. Visualization tasks, including training curves and confusion matrices, were implemented using Matplotlib, Seaborn, and psutil, along with OS and time, which were used for runtime and resource utilization monitoring.

All experiments were executed in Google Colab Pro, leveraging its cloud-based infrastructure with 51 GB RAM and GPU acceleration (NVIDIA Tesla T4/A100) to efficiently train and evaluate the deep learning models. Collectively, the integration of the Cuckoo Sandbox for dynamic behavior monitoring with Python-based libraries and tools for feature extraction, data transformation, and model training established a robust end-to-end experimental pipeline for ransomware classification, evaluation, and result visualization.

## 4.2 Evaluation metrics

Evaluation metrics are critical for objectively quantifying the performance of machine learning models and providing standardized measures to compare models and assess their predictive effectiveness. These metrics provide insights into how well a model generalizes to unseen data and its ability to make accurate predictions. The primary tool employed in this study was a confusion matrix, which systematically evaluates the classification performance by comparing the predicted labels with the ground-truth labels. The confusion matrix consists of four elements: True Positives (TP), True Negatives (TN), False Positives (FP), and False Negatives (FN). TP and TN correspond to correctly identified ransomware and benign samples, respectively, while FP represents benign samples incorrectly classified as ransomware and FN indicates ransomware samples misclassified as benign. This detailed structure provides a clear visualization of the correct and incorrect classifications, serving as the foundation for deriving the performance metrics. From this matrix, core metrics including accuracy, precision, recall, F1-score, and area under the ROC curve (AUC) were computed to comprehensively assess model performance. These metrics collectively capture overall correctness, detection sensitivity, misclassification trade-offs, and threshold-independent discrimination. Their formal definitions, descriptions, and equations are presented in **Table 5**.

**Table 5** Evaluation metrics of the proposed approach.

| Metric Name | Description | Related Syntax/Equation |
|---|---|---|
| Accuracy | The ratio of correctly predicted instances (TP and TN) to the total number of observations. | $Accuracy = \frac{TP+TN}{TP+TN+FP+FN}$ |
| Precision | The ratio of correctly predicted positive observations (TP) to the total predicted positives (TP + FP). | $Precision = \frac{TP}{TP+FP}$ |
| Recall (Sensitivity) | The proportion of actual positives (TP) correctly identified by the model. | $Recall = \frac{TP}{TP+FN}$ |
| F1-Score | The harmonic mean of precision and recall, balancing false positives and false negatives. | $F1\text{-}measure = 2 \times \frac{Precision \times recall}{Precision + recall}$ |
| AUC (Area Under ROC Curve) | Evaluates classifier performance across thresholds; higher values indicate stronger separability. | Derived from ROC curve (TPR vs. FPR). |

In the context of ransomware detection, these metrics are particularly significant, as minimizing false negatives is critical to prevent undetected ransomware infections, while reducing false positives helps avoid benign software misclassification, thereby enhancing the practical reliability of the proposed approach in real-world deployments

## 4.3 Dataset splitting and evaluation strategy

The final corpus consists of 1,000 behavior-derived color images (500 ransomware, 500 benign), each generated from the 100 dynamic behavioral features described in Section 3 and transformed into a fixed-size visual representation suitable for CNN backbone processing.

Model evaluation was conducted using a stratified 5-fold cross-validation (CV) protocol on the entire dataset to obtain an unbiased estimate of the generalization performance. In each fold, the dataset was partitioned such that 80% of the samples (four folds) were used exclusively for training, and the remaining 20% (one fold) served as the hold-out validation set. Class stratification preserved the 1:1 ransomware–benign ratio in both training and validation splits. This process was repeated five times so that each sample appeared in the validation set exactly once, and in the training set, four times across folds. All reported metrics represent the mean ± standard deviation of the five validation runs.

In each cross-validation fold, the classifier is trained on the fold's training split with per-sample weights determined by an RL agent. The agent assigns these weights prior to model fitting, guiding the classifier to focus on informative samples. Evaluation was performed on the corresponding test split, and the mean ± standard deviation of the accuracy, precision, recall, F1-score, and AUC were reported across folds. Predictions from all folds were combined to produce aggregate confusion matrices and ROC curves, providing a clear overview of the overall classification performance.

*4.4 Experimental evaluation*

This section presents a comprehensive evaluation of the proposed TL-RL-FusionNet, which integrates dual-CNN feature extraction with a lightweight RL-guided residual MLP classifier for efficient decision-making. The framework was benchmarked against six baseline classifiers: RF, XGBoost, SVM, LR, CNN, and ANN, each evaluated with and without RL-based sample weighting.

The evaluation followed a multi-step process.
1. Cross-validation performance and visualization of classification metrics via confusion matrices and ROC curves
2. Computational efficiency analysis in terms of execution time and memory usage.
3. Post-hoc model interpretability analysis using t-SNE and Grad-CAM visualizations
4. Comparison with related studies to contextualize the results.

The findings, supported by detailed figures and tables, demonstrate the robustness, efficiency, and practical applicability of TL-RL-FusionNet for ransomware detection. Notably, the RL-enhanced residual MLP achieved the highest performance, surpassing all baseline models and their RL-free counterparts.

*4.4.1 Cross-validation performance*

This subsection evaluates all classifiers under the RL-guided sample weighting and baseline configurations. The analysis combined statistical metrics, confusion matrix patterns, and ROC curve assessments. This evaluation provides a comprehensive view of the classification behavior of each model by comparing the detection accuracy, error distribution, and sensitivity-specificity trade-offs in ransomware detection.

**Classification performance of all classifiers, with and without RL-guided sample weighting:** The cross-validation results for all classifiers, evaluated with and without RL-guided sample weighting, are summarized in **Table 6**, while the corresponding relative accuracy improvements are visualized **Fig. 4**. The performance metrics, reported as mean ± standard deviation over five folds, included accuracy, precision, recall, F1-score, and AUC. This study used accuracy as the primary evaluation metric because of its central role in assessing ransomware detection performance, whereas the other metrics were reported to provide a more complete evaluation. The results demonstrate that the RL-guided sample weighting consistently enhances the performance of most classifiers. The proposed RL-enhanced residual MLP achieved the highest overall performance, attaining an accuracy of 99.10% and an F1-score of 99.11%, outperforming its non-RL counterpart (98.80% accuracy, 98.81% F1-score). RL integration also improved recall (99.60% vs. 99.20%) and AUC (99.74% vs. 99.72%), indicating superior sensitivity and stronger class separability. Similar gains were observed for CNN, ANN, RF, and XGBoost, whereas SVM exhibited negligible improvement, likely because of its limited adaptability to dynamic sample weighting in high-dimensional spaces.

Table 6 Cross-validation performance of all classifiers with and without RL-guided sample weighting, showing mean ± standard deviation for accuracy, precision, recall, F1-score, and AU

| Model | RL/Without RL | Accuracy | Precision | Recall | F1 Score | AUC |
| --- | --- | --- | --- | --- | --- | --- |
| **MLP** | With RL | 0.9910 ± 0.0108 | 0.9863 ± 0.0163 | 0.9960 ± 0.0055 | 0.9911 ± 0.0107 | 0.9974 ± 0.0042 |
|  | Without RL | 0.9880 ± 0.0045 | 0.9843 ± 0.0111 | 0.9920 ± 0.0084 | 0.9881 ± 0.0044 | 0.9972 ± 0.0030 |
| **CNN** | With RL | 0.984 ± 0.0082 | 0.9803 ± 0.0137 | 0.988 ± 0.0130 | 0.9841 ± 0.0082 | 0.9975 ± 0.0020 |
|  | Without RL | 0.983 ± 0.0045 | 0.9842 ± 0.0112 | 0.982 ± 0.0130 | 0.9830 ± 0.0045 | 0.9979 ± 0.0019 |
| **ANN** | With RL | 0.9820 ± 0.0084 | 0.9764 ± 0.0109 | 0.9880 ± 0.0110 | 0.9821 ± 0.0083 | 0.9950 ± 0.0049 |
|  | Without RL | 0.9810 ± 0.0108 | 0.9728 ± 0.0185 | 0.9900 ± 0.0100 | 0.9812 ± 0.0107 | 0.9955 ± 0.0043 |
| **LR** | With RL | 0.9800 ± 0.0094 | 0.9744 ± 0.0111 | 0.986 ± 0.0134 | 0.9801 ± 0.0093 | 0.9948 ± 0.0050 |
|  | Without RL | 0.9790 ± 0.0082 | 0.9744 ± 0.0111 | 0.9840 ± 0.0114 | 0.9791 ± 0.0082 | 0.9946 ± 0.0039 |
| **XGBoost** | With RL | 0.9780 ± 0.0057 | 0.9784 ± 0.0185 | 0.9780 ± 0.0084 | 0.9781 ± 0.0054 | 0.9962 ± 0.0034 |
|  | Without RL | 0.9760 ± 0.0065 | 0.9708 ± 0.0189 | 0.9820 ± 0.0130 | 0.9762 ± 0.0063 | 0.9961 ± 0.0033 |
| **RF** | With RL | 0.9770 ± 0.0084 | 0.9651 ± 0.0155 | 0.9900 ± 0.0100 | 0.9773 ± 0.0081 | 0.9957 ± 0.0044 |
|  | Without RL | 0.974 ± 0.0082 | 0.9650 ± 0.0155 | 0.984 ± 0.0114 | 0.9743 ± 0.0081 | 0.9956 ± 0.0038 |
| **SVM** | With RL | 0.9660 ± 0.0055 | 0.9534 ± 0.0075 | 0.9800 ± 0.0122 | 0.9665 ± 0.0055 | 0.9888 ± 0.0092 |
|  | Without RL | 0.9660 ± 0.0055 | 0.9534 ± 0.0075 | 0.9800 ± 0.0122 | 0.9665 ± 0.0055 | 0.9888 ± 0.0092 |

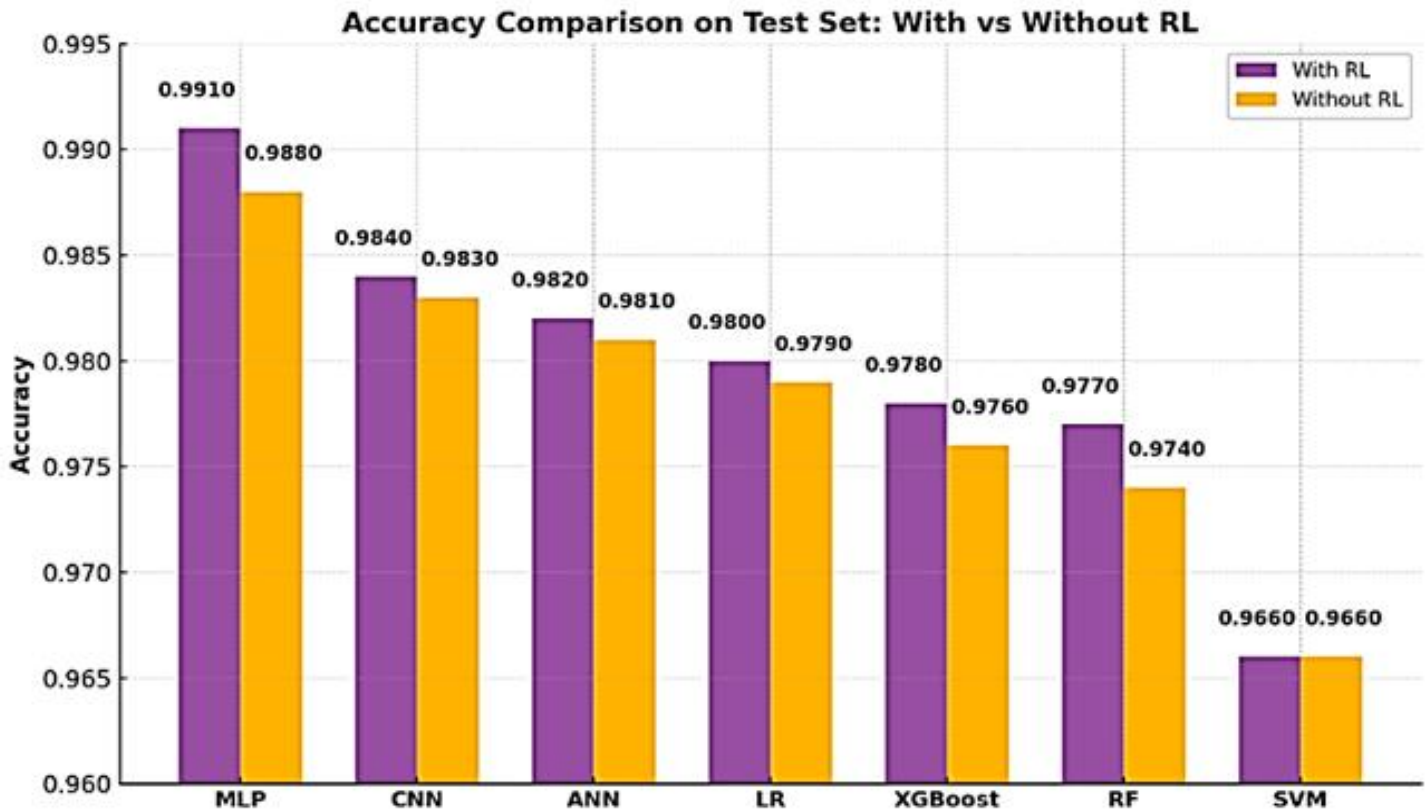

**Fig. 4.** Accuracy comparison of models with and without RL-guided sample weighting. RL-enhanced models show improvements across classifiers, with proposed MLP achieving best performance. The comparison demonstrates RL-guided adaptive sample weighting's effectiveness in prioritizing hard-to-classify instances, reducing overfitting and enhancing generalization. In ransomware detection, improved recall is crucial as it reduces false negatives and undetected threats.

To better understand the classification behavior of the evaluated models, fold-aggregated confusion matrices were generated for each baseline and RL-guided variant (**Table 7**). Each matrix reports the counts of true positives (TP), true negatives (TN), false positives (FP), and false negatives (FN) for ransomware and benign classes. A comparative inspection revealed that the RL-guided models consistently reduced the number of false negatives across most classifiers while maintaining similar or lower false-positive counts. This suggests that the RL weighting improves decision calibration in borderline classification cases, leading to more robust detection. For example, both the RL-augmented CNN and ANN demonstrate a notable reduction in FN instances compared to their non-RL counterparts, which is particularly critical in ransomware detection, where missed threats pose severe risks.

The most significant improvement was observed in the proposed RL-MLP model, which achieved the lowest misclassification rate across all the tested models. Specifically, it recorded only three ransomware instances misclassified as benign (FN) and six benign instances misclassified as ransomware (FP), yielding an overall accuracy of 99.10%. In contrast, its non-RL counterpart recorded four FN and eight FP cases, corresponding to a slightly lower accuracy of 98.80%. These results highlight the ability of RL-guided weighting to emphasize hard-to-classify samples, thereby strengthening the discriminative boundary between ransomware and benign samples.

**Threshold-based discriminative performance using ROC curves:** To further evaluate the threshold-based discrimination capability, Receiver Operating Characteristic (ROC) curves were generated for all models. As shown in **Fig. 5**, the ROC analysis, aggregated over all five cross-validation folds, provides a comprehensive view of the model discrimination performance across diverse data splits. By pooling fold-wise predictions, we ensured that the computed AUC values captured the average behavior of the model rather than being biased by any single fold. The proposed TL-RL-FusionNet framework, particularly the RL-weighted MLP, CNN, and tree-based ensembles, demonstrated an AUC of $\approx 1.00$, indicating almost perfect separability between ransomware and benign samples. The tight clustering of the ROC curves near the top-left corner reinforces the models' ability to maintain high sensitivity with negligible false positives, which is critical for operational ransomware detection. Furthermore, the consistency of these results with the unseen test set evaluation (AUC $\approx 0.99$) underscores the reliability and generalization capacity of the proposed framework. This confirms that the integration of reinforcement learning-driven sample weighting with dual-CNN feature fusion not only improves the classification accuracy but also stabilizes the performance across folds, reducing the variance and overfitting risk.

**Table 7** Aggregated confusion matrices for all classifiers with and without RL-guided sample weighting, illustrating the distribution of TP, TN, FP, and FN counts. The RL-guided MLP configuration achieved the most optimal results, underscoring the benefit of dynamic weighting through reinforcement learning.

| Model Name | Confusion matrix with RL | Confusion matrix without RL |
|---|---|---|
| MLP (Proposed) | TN 493, FP 6, FN 3, TP 498 | TN 492, FP 8, FN 3, TP 497 |
| CNN | TN 490, FP 10, FN 5, TP 495 | TN 492, FP 8, FN 4, TP 496 |
| ANN | TN 488, FP 12, FN 6, TP 494 | TN 486, FP 14, FN 5, TP 495 |
| LR | TN 487, FP 13, FN 7, TP 493 | TN 487, FP 13, FN 8, TP 492 |
| XGBoost | TN 489, FP 11, FN 5, TP 495 | TN 485, FP 15, FN 9, TP 491 |
| RF | TN 482, FP 18, FN 5, TP 495 | TN 482, FP 18, FN 6, TP 494 |
| SVM | TN 476, FP 24, FN 10, TP 490 | TN 476, FP 24, FN 10, TP 490 |

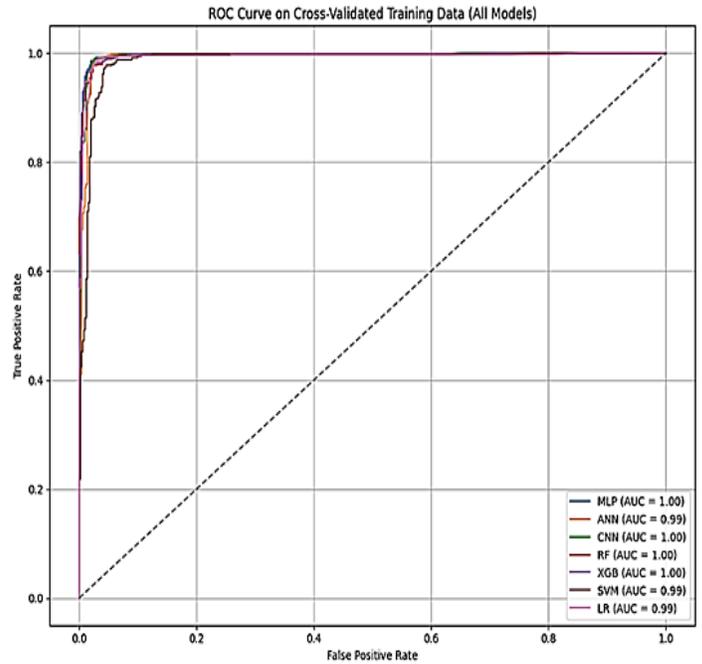

**Fig. 5.** ROC curves for all models were evaluated using cross-validated training data. The curves illustrate the trade-off between the true and false positive rates across the classification thresholds. MLP, CNN, RF, and XGBoost achieved perfect separability (AUC = 1.00), while ANN, SVM, and LR followed closely with AUC = 0.99. The results confirmed the high sensitivity and specificity of the model, particularly those augmented by reinforcement learning.

*4.4.2 Computational efficiency evaluation*

In addition to predictive accuracy, computational efficiency is an essential requirement for ransomware detection systems, especially in environments with limited resources or those that require real-time processing. Therefore, we evaluated the efficiency in terms of training time, inference time, and memory usage (RAM) across all classifiers under both RL-guided and non-RL training schemes. The results are summarized in **Table 8** and visualized in **Fig. 6**. and **Fig. 7**.

**Timing performance analysis (training and inference time):** Across nearly all models, RL-guided sample weighting resulted in shorter training times and faster inference than non-RL counterparts, as illustrated in Fig. 6 (a) and 6 (b). For instance, the training time for CNN decreased from 76.96 s to 25.49 s, representing a 67% reduction. Similarly, the MLP and ANN exhibited significant reductions in training time, from 16.70 s to 13.78 s for MLP, and from 4.97 s to 4.15 s for ANN. As shown in Fig. 9, the inference time differences between the RL and non-RL variants were relatively minor across all models. The proposed MLP achieved an inference time of 0.23 s with RL, compared to 0.36 s without RL, and the CNN inference time decreased from 0.28 s to 0.164 s. Lightweight classifiers, such as XGBoost, RF, and LR, maintained inference times below 0.04 s, affirming their appropriateness for low-latency applications.

**Table 8** Computational efficiency comparison of all models with and without RL-guided sample weighting, including average training time, inference time, and memory usage (RAM) across five-folds.

| Model | RL/Without RL | Train Time (s) | Inference Time (s) | Train RAM (MB) | Inference RAM (MB) |
|---|---|---|---|---|---|
| **MLP** | With RL | 13.78 ± 0.13 | 0.23 ± 0.01 | 73.80 ± 48.36 | 0.66 ± 1.44 |
|  | Without RL | 16.70 ± 0.38 | 0.36 ± 0.03 | 89.55 ± 70.97 | 2.73 ± 1.81 |
| **CNN** | With RL | 25.494 ± 0.4676 | 0.164 ± 0.0055 | 163.488 ± 67.0153 | 6.21 ± 5.59 |
|  | Without RL | 76.960 ± 1.3843 | 0.280 ± 0.0224 | 223.318 ± 75.6966 | 15.258 ± 12.7609 |
| **ANN** | With RL | 4.15 ± 0.10 | 0.11 ± 0.01 | 17.30 ± 23.44 | 3.16 ± 2.10 |
|  | Without RL | 4.97 ± 0.12 | 0.16 ± 0.03 | 44.56 ± 49.24 | 3.24 ± 2.15 |
| LR | With RL | 3.044 ± 0.4058 | 0.034 ± 0.0114 | 0.156 ± 0.1424 | 0.000 ± 0.0000 |
|  | Without RL | 3.95 ± 0.49 | 0.04 ± 0.01 | 0.55 ± 0.67 | 0.00 ± 0.00 |
| XGBoost | With RL | 3.71 ± 0.63 | 0.01 ± 0.00 | 12.39 ± 25.84 | 0.000 ± 0.0000 |
|  | Without RL | 4.696 ± 0.6352 | 0.010 ± 0.0000 | 18.488 ± 35.2322 | 0.050 ± 0.1118 |
| RF | With RL | 1.75 ± 0.08 | 0.01 ± 0.00 | 0.05 ± 0.12 | 0.00 ± 0.00 |
|  | Without RL | 2.280 ± 0.0800 | 0.020 ± 0.0000 | 0.088 ± 0.1968 | 0.000 ± 0.0000 |
| SVM | With RL | 1.80 ± 0.06 | 0.24 ± 0.01 | 6.17 ± 13.36 | 0.02 ± 0.05 |
|  | Without RL | 2.07 ± 0.03 | 0.31 ± 0.01 | 7.03 ± 10.92 | 0.51 ± 1.14 |

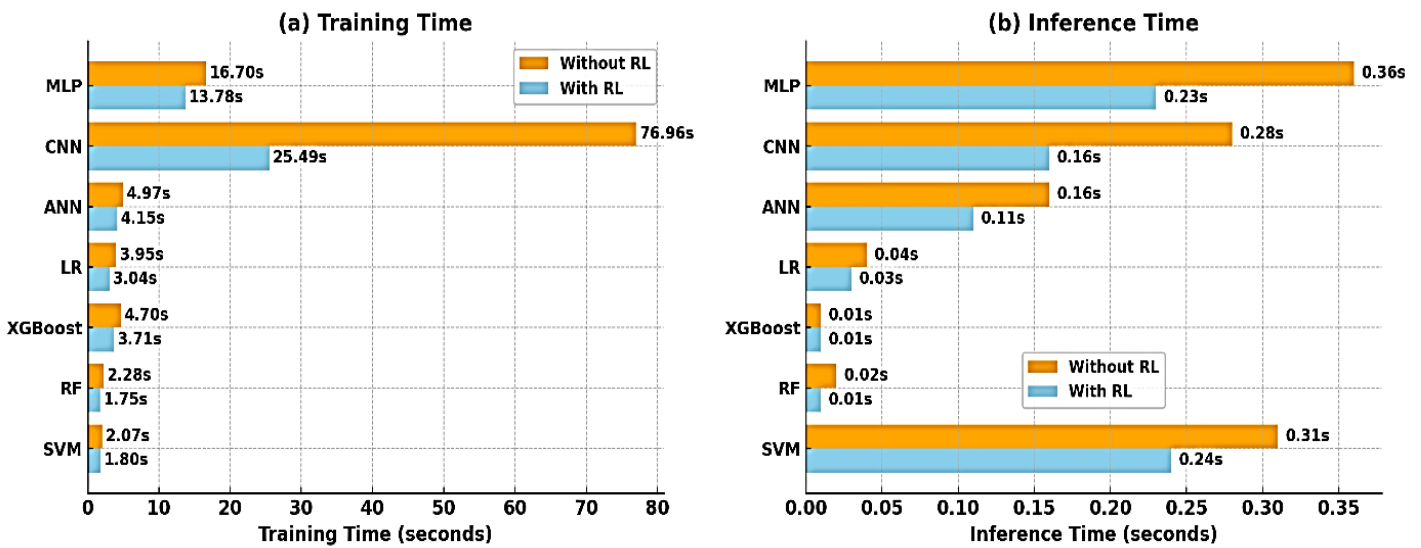

**Fig. 6.** Training and inference time comparison of ransomware detection models with and without RL-guided sample weighting. (a) Training time in seconds and (b) inference time in seconds across seven classifiers (SVM, RF, XGBoost, LR, ANN, CNN, and MLP). RL integration significantly accelerates convergence in deep models (e.g., CNN and MLP) while maintaining or improving inference efficiency across all models.

**Memory usage**
Fig. 7 (a) and Fig. 7 (b) show the RAM usage during the training and inference phases, respectively. During training, the application of RL-guided weighting significantly reduced the memory overhead by preventing redundant updates, as evidenced in the cases of CNN (223 MB reduced to 163 MB) and ANN (44 MB reduced to 17 MB). In the inference phase, the RL-weighted MLP reduces memory consumption from 2.73 MB (non-RL) to 0.66 MB (RL), whereas CNN shows the largest drop, from 15.26 to 6.21 MB. The ANN usage declines slightly from 3.24 to 3.16 MB, XGBoost maintains a negligible footprint (0.05 to 0.00 MB), and SVM decreases from 0.51 to 0.02 MB. LR and RF require no additional memory in either configuration (0.00 MB). Although some models exhibit lower memory usage, their accuracies are significantly lower than those of the optimized MLP. The RL-weighted MLP offers the best trade-off, combining high accuracy with low RAM inference consumption.

These results highlight an important finding: rather than incurring additional computational overhead, our lightweight tabular Q-learning agent improves both efficiency and accuracy through its sample reweighting mechanism. During training, the agent down-weights easy samples, such as benign applications that perform only basic file I/O operations or ransomware families with obvious, repetitive encryption behaviors that are consistently classified with high confidence. By emphasizing harder or misclassified cases, the agent reduces redundant gradient updates on easy samples, effectively lowering the number of costly backpropagation operations without reducing the dataset size. Consequently, the models converge faster and require fewer resources during training.

At inference, no reinforcement learning (RL) component is executed. The improved efficiency observed here is therefore an indirect effect of better-calibrated models trained with RL weighting: predictions are more stable and memory allocation is more efficient, leading to faster and lighter inference.

Overall, the RL-guided sample weighting not only enhances the detection robustness but also improves the computational efficiency. By streamlining the training dynamics and stabilizing the inference behavior, the lightweight Q-learning agent enables faster convergence, lower memory usage, and reduced computational cost without introducing overhead. These findings underscore the novelty of TL-RL-FusionNet and reinforce its deployability in scenarios where both high accuracy and resource efficiency are essential.

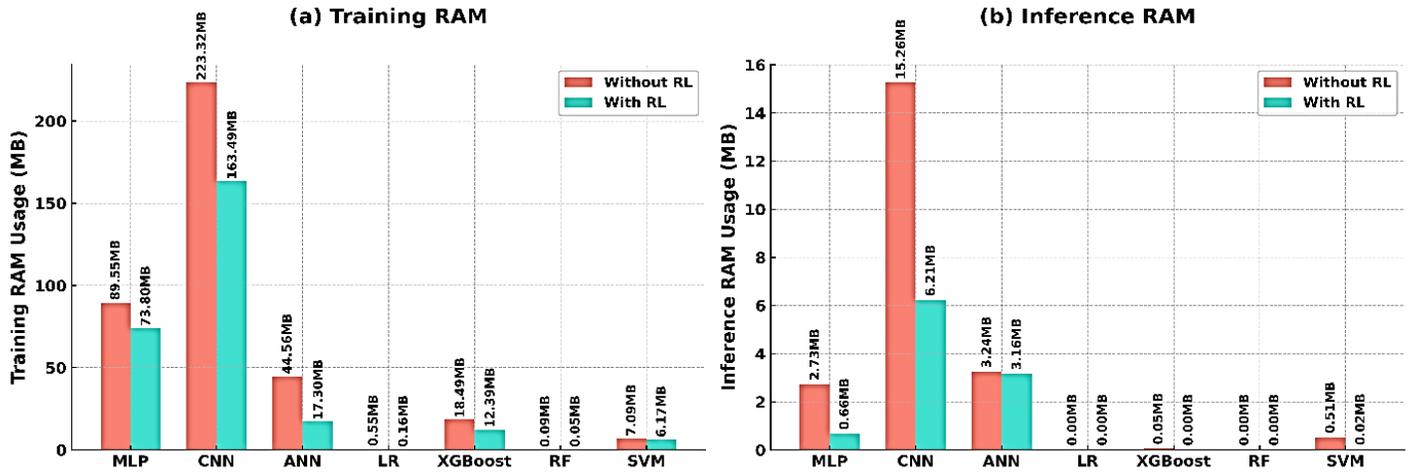

**Fig. 7.** RAM usage comparison of ransomware detection models with and without RL-guided sample weighting. (a) Training RAM usage and (b) inference RAM usage in megabytes across seven classifiers. RL-guided models consistently reduce memory overhead, with the most pronounced efficiency gains observed in CNN and MLP, demonstrating the framework's suitability for resource-constrained or real-time environments.

*4.4.3 Model interpretability analysis*

To ensure that the model predictions are not only accurate but also explainable, a series of interpretability techniques were applied to the RL-enhanced classifiers. This section explores two key aspects of model performance: the distribution of learned features and the explainability of its decision-making process. First, we visualized the learned feature space using t-SNE to assess how well the model differentiates between ransomware and benign samples. Second, we employed Gradient-weighted Class Activation Mapping (Grad-CAM) to provide insights into the specific image regions that influence the model's decisions. By analyzing these aspects, we aim to provide a comprehensive view of both the global feature distribution and localized decision cues, enhancing our understanding of the model's robust and discriminative behavior on unseen data.

**Feature space visualization and misclassified analysis using t-SNE:** To gain deeper insights into the internal representation of the learned features, t-distributed Stochastic Neighbor Embedding (t-SNE) visualization was applied to the final test set features.

As illustrated in **Fig. 8**, the two-dimensional projection reveals a clear separation between the ransomware (R) and benign (B) classes. Although some samples appear near the decision boundary, the majority form distinct and compact clusters, indicating the effectiveness of the feature extraction and learning pipeline in capturing class-specific representations. This implies that the proposed RL-enhanced MLP model is capable of learning well-generalized decision boundaries. The ransomware samples (green) tended to aggregate into distinct regions, whereas the benign samples (orange) formed compact, isolated clusters. This visual separation supports the model's effectiveness in capturing high-level discriminative patterns from behavior-transformed image features, enhancing classification confidence and interpretability.

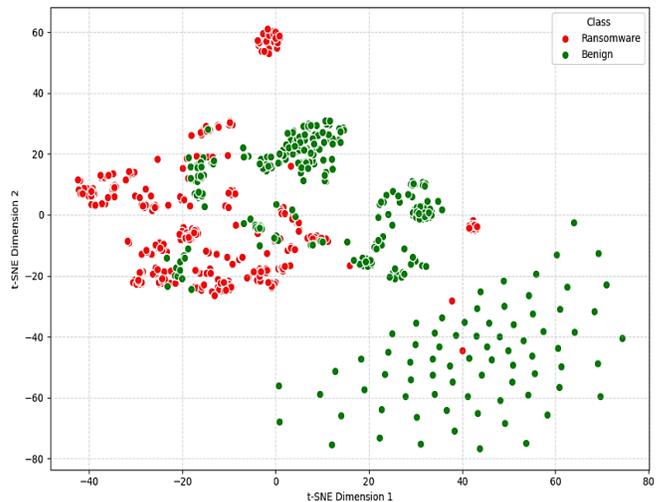

**Fig. 8.** t-SNE plot of the final test set features. This plot shows a 2D projection of the learned features. Green and orange points represent ransomware (R) and benign (B) samples, respectively. The formation of distinct clusters indicates strong feature separability with minimal class overlap, reinforcing the model's ability to learn discriminative representations for accurate classification.

To further investigate the model decision boundaries, we applied t-distributed Stochastic Neighbor Embedding (t-SNE)

to project high-dimensional feature representations into two- and three-dimensional spaces (**Fig. 9**). In both projections, the correctly classified samples (blue) formed distinct clusters, reflecting the model's ability to effectively separate classes. Misclassified benign samples (green, FP) and ransomware samples (red, FN) tended to cluster near the decision boundary, indicating that these cases shared feature similarities with the opposite class. The 3D visualization further highlights that misclassifications occur in overlapping regions where class boundaries are less distinct. Overall, the combined confusion matrix and t-SNE analyses confirmed that the RL-guided adaptive weighting enhanced the reliability of ransomware detection by minimizing high-impact false negatives and improving separation in ambiguous regions of the feature space. These findings demonstrate the robustness of the RL-MLP in handling adversarially evasive ransomware behaviors while maintaining generalization to benign software.

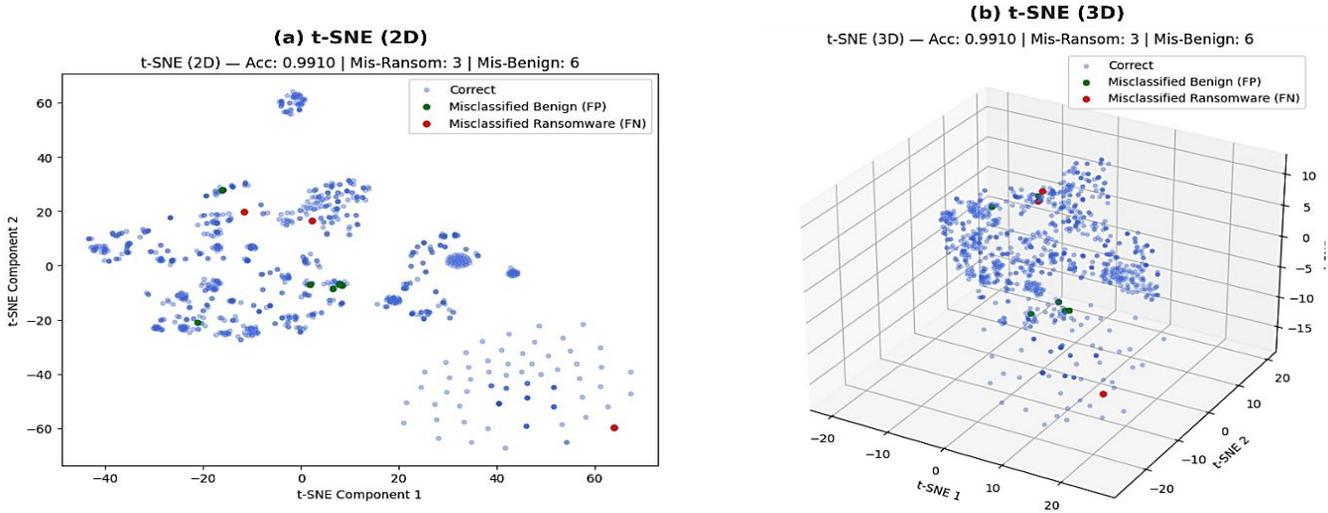

**Fig. 9.** Visualization of misclassified samples for the RL-MLP model using t-SNE. (a) 2D t-SNE projection and (b) 3D t-SNE projection. The correctly classified samples appear in blue, false positives (benign misclassified as ransomware) appear in green, and false negatives (ransomware misclassified as benign) appear in red. The figures show that samples form distinct clusters, whereas misclassified instances lie near decision boundaries, highlighting the inherent ambiguity of these samples.

**Saliency map analysis using Grad-CAM**

To improve interpretability, we employed Gradient-weighted Class Activation Mapping (Grad-CAM) to visualize which parts of the feature-to-image representations influenced model predictions. Grad-CAM is important because it makes the model's "reasoning" more transparent, showing whether decisions are based on meaningful behavioral patterns rather than random noise.

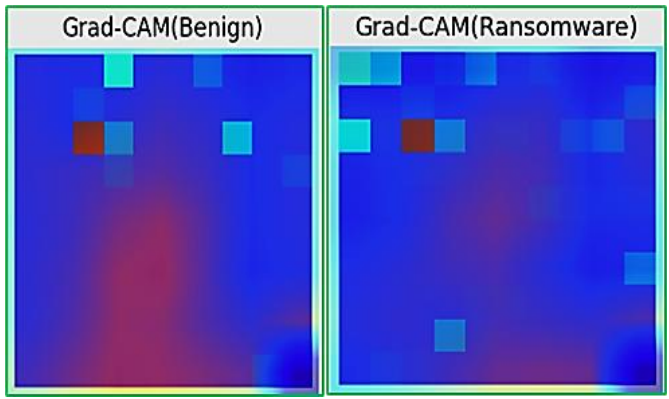

**Fig. 10.** Grad-CAM heatmaps show visual saliency maps for benign and ransomware inputs, highlighting influential classification regions. RGB image regions correspond to behavioral features (file operations, registry edits, and network activity) with distinct activation patterns. Ransomware heatmaps show hotspots in file and registry activities, whereas benign samples display diffuse patterns, confirming the model's focus on ransomware behavior.

As shown in **Fig. 10,** the benign sample displayed diffuse activations, suggesting no single dominant feature group, whereas the ransomware sample exhibited more localized hotspots. Because each image region corresponds to feature categories such as file operations, registry edits, or network activity, these focused activations indicate that the model relies on ransomware-relevant behaviors. This is particularly important for detecting polymorphic and mutating ransomware, where surface-level code changes vary, but the underlying behaviors (e.g., file encryption and registry modification) remain consistent.

Thus, Grad-CAM not only validates that our model learns the correct cues but also strengthens trust in its ability to generalize across diverse ransomware families.

*4.4.4 Comparison with existing relevant methods*

To further evaluate the effectiveness of the proposed TL-RL-FusionNet framework, we benchmarked its performance against ransomware detection techniques from the literature that also combined pretrained CNN feature extractors with separate classifiers. The comparative results are presented in **Table 9**. It is important to note that the studies listed in Table 6 differ substantially in terms of datasets, feature engineering approaches (e.g., static, dynamic, or hybrid analysis), and targeted malware families or versions. Consequently, the results should be interpreted as contextual benchmarks across related methodologies, rather than a direct head-to-head evaluation.

Our method is based on a custom, balanced, threat-informed dynamic behavioral dataset, whereas most prior studies rely primarily on static binary features, memory dumps, and hybrid data sources.

**Table 9** Comparison of the proposed TL-RL-FusionNet framework with existing ransomware detection methods employing pretrained CNN feature extractors paired with separate classifiers. The results highlight the superior performance of the proposed RL-enhanced Residual Lightweight MLP classifier, which combines dynamic-to-image feature transformation with RL-guided sample weighting.

| Ref. | Analysis Technique | Feature Extractor(s) | RL-based Sample Weighting | Classifier | Accuracy (%) | Training Time (s) | Prediction Time (s) | Memory Usage (MB) | False Positive (%) | False Negative (%) |
|---|---|---|---|---|---|---|---|---|---|---|
| [14] | Static | DenseNet-121 | No | XGBoost | 98.58 | – | – | – | – | – |
| [15] | Static | VGG-16 + ResNet50 | No | Bi-KNN | 99.00 | 18.928 | – | – | – | – |
| [13] | Static | VGG16, VGG19, ResNet50, InceptionV3 | No | MLP | 98.55 | – | 0.471 | – | – | – |
| [16] | Static | VGG-16 | No | SVM | 92.97 | – | – | – | – | – |
| [21] | Static | - | No | DRL (A2C, DQN, DDQN, PPO) | 89.78 with A2C | 168.27 | 0.76 | | | |
| [31] | Dynamic | - | - | FL and RL | Above 98% | - | - | - | | |
| Our method (TL-RL-FusionNet) | Dynamic | EfficientNetB0 + InceptionV3 | Yes | RL-enhanced Residual MLP | 99.10 | 13.90 | 0.23 | 0.66 | 0.70 | 0.20 |

Furthermore, while previous studies have demonstrated competitive accuracy, such as 99% with VGG-16 + ResNet50 using Bi-KNN [15] or 98.58% with DenseNet-121 and XGBoost [14], none have incorporated a RL-based sample weighting mechanism. In contrast, the study in [21] introduced Deep Reinforcement Learning (DRL) models such as A2C, DQN, DDQN, and PPO for static ransomware detection; however, their best performance (A2C = 89.78%) was notably lower, with long training durations (168 s) and moderate inference latency (0.76 s). The more recent CyberForce framework [31] extended RL into a federated dynamic environment, integrating FL with RL for decentralized malware mitigation and achieving above 98% accuracy. Nevertheless, CyberForce primarily targets adaptive defense strategies in IoT networks rather than direct ransomware classification.

In comparison, our proposed TL-RL-FusionNet, incorporating an RL-enhanced Residual Lightweight MLP classifier, achieved 99.1% accuracy, a 0.70% false-positive rate, and a 0.20% false-negative rate, with an average inference memory footprint of only 0.66 MB and prediction latency of 0.156 s. To the best of our knowledge, no prior study has employed a dynamic-to-image transformation pipeline combined with RL-guided transfer learning for ransomware detection, making our contribution novel in terms of both methodology and performance.

Moreover, none of the compared studies reported both false-positive and false-negative rates, which are highly relevant in cybersecurity contexts. False negatives are particularly critical because they correspond to undetected threats that can bypass the defenses. The extremely low false-negative rate achieved by TL-RL-FusionNet underscores its operational reliability. In addition, the compared studies do not address computational efficiency in terms of inference memory usage and do not perform model interpretability analyses. Our framework addresses these gaps through t-SNE-based visualization of learned feature spaces and Grad-CAM saliency mapping to identify decision-critical regions for both benign and ransomware samples.

TL-RL-FusionNet outperforms existing approaches in terms of accuracy and robustness while providing computational efficiency and interpretability, making it deployable for ransomware detection in IoT and edge devices.

## 5. Discussion

This section discusses the experimental findings of TL-RL-FusionNet, highlighting the importance of reinforcement learning, its influence on the results, and the advancements of the framework over prior studies. It then outlines the scientific and practical implications, acknowledges key limitations with future directions, and ends with a takeaway underscoring RL as the defining factor that makes TL-RL-FusionNet adaptive and deployable.

Ransomware is increasingly engineered to evade detection by altering behaviors or mimicking benign activity. In this setting, static weighting of training samples is insufficient because it treats all data points as equally informative. The central contribution of this study is demonstrating that reinforcement learning (RL) directly addresses this limitation by adaptively reweighting training samples.

The RL agent enhanced the framework by dynamically reweighting samples: misclassified or challenging cases were assigned higher weights, whereas trivial, consistently correct cases were down-weighted. This dynamic prioritization improved the classifier's ability to learn from deceptive behaviors such as registry edits, subtle file-system operations, and suspicious network traffic. Although RL operated at the sample level rather than individual features, the indirect effect was that these high-risk behavioral categories received greater influence during training. This mechanism improves the calibration around difficult decision boundaries and reduces high-impact errors. As shown in the fold-aggregated confusion matrices (Table 7), the RL-guided MLP achieved only three false negatives and six false positives (compared with four FN and eight FP without RL). This corresponds to a measurable reduction in the error rates (0.30% FN, 0.70% FP), which are critical for reliable ransomware detection. Moreover, by consistently highlighting misclassified or borderline cases, the RL agent improved its resilience against polymorphism and feature obfuscation. Simultaneously, interpretability analyses further reinforced operational trust: t-distributed stochastic neighbor embedding projections showed clear class separation and identified misclassified samples, gradient-weighted class activation mapping highlighted semantically meaningful behavioral cues, and confusion matrices documented low FN rates, underscoring the framework's reliability in minimizing undetected threats. Together, these results highlight the dual value of TL-RL-FusionNet: it is both effective and trustworthy.

Previous research on RL in malware detection has largely focused on static features such as portable executable headers [19], Android permissions [21], and Bayesian vulnerability modelling [20] or relied on pretrained CNNs without adaptive weighting or efficiency profiling [14], [15], [16]. Additionally, our previous work [18] employed full end-to-end fine-tuning of multiple CNN backbones (ResNet50, EfficientNetB0, InceptionV3, Xception, and VGG), achieving near-perfect accuracy (99.96% with ResNet50) on color-transformed dynamic features. However, this approach incurs substantial computational costs owing to backpropagation across millions of parameters, resulting in prolonged training times and high memory consumption. TL-RL-FusionNet advances the field by freezing EfficientNetB0 and InceptionV3 as feature extractors and introducing RL-guided sample weighting with a residual MLP classifier. This approach retains an accuracy above 98–99% while reducing the training time by approximately 45–55% and memory use by approximately 70–80%. In doing so, it directly addresses one of the key weaknesses of our earlier work: computational efficiency. Compared to the baseline models without RL weighting, TL-RL-FusionNet achieved consistently higher performance across accuracy, precision, recall, and AUC. Classical machine learning models such as RF, SVM, and XGBoost showed modest gains, but the largest improvements appeared in deep learning models, where RL weighting stabilized training and sharpened decision boundaries.

The findings show that reinforcement learning can serve as a practical mechanism for guiding deep learning models in security contexts. More broadly, the principle of adaptive sample weighting can be applied across domains where difficult edge cases carry the greatest risks, including fraud analysis and healthcare diagnostics. In addition, by combining frozen pretrained backbones with RL-guided adaptivity, TL-RL-FusionNet achieves a rare balance of accuracy, efficiency, and interpretability. This efficiency has direct practical implications: the framework can operate effectively in resource-constrained environments such as IoT devices, edge systems, or mobile platforms, where computational and memory budgets are limited but robust detection remains essential.

Despite its contributions, the TL-RL-FusionNet has some limitations. First, although the RL weighting improved robustness, it was not a complete defense. Persistently misclassified poisoning samples may still be repeatedly up weighted, reinforcing incorrect decision boundaries, and adversarial behaviors carefully engineered to mimic benign activity could still bypass detection. Moreover, adversarial robustness was not explicitly tested in this study, representing a promising direction for future validation under adversarially controlled environments Second, the evaluation was conducted on a balanced dataset of 1,000 samples curated from recent ransomware families. Although this mitigated class imbalance and ensured representativeness, larger and more diverse datasets are required to comprehensively assess scalability and robustness. Although representative, larger, and more diverse datasets are required to confirm the scalability across families and execution environments. Third, the reinforcement learning component employs a tabular Q-learning agent with discrete weight actions. Although effective in this context, such agents may be less adaptable in complex feature spaces than more advanced RL methods, limiting their generalization potential. These limitations open clear avenues for future research.

In summary, reinforcement learning is the key to making TL-RL-FusionNet adaptive. It shifts the training process away from static, one-size-fits-all learning and toward a targeted focus on the cases most likely to evade detection. This ability to combine high accuracy, low error rates, and practical efficiency positions TL-RL-FusionNet as a credible foundation for next-generation ransomware detection and as a steppingstone for future advances in adaptive security AI.

## 6. Conclusion and future work

This study demonstrated that the evolving behavior of ransomware requires adaptive detection strategies beyond static weighting. TL-RL-FusionNet addressed this by using reinforcement learning to reweight challenging samples, such as subtle file-system operations, registry edits, and suspicious network traffic, thereby reducing critical misclassifications, improving resilience against evasion, and maintaining high accuracy, efficiency, and interpretability. By integrating frozen dual CNN backbones (EfficientNetB0 and InceptionV3) with a residual MLP classifier and an RL-based weighting agent, TL-RL-FusionNet achieved a principled balance between the predictive performance and computational efficiency. Across stratified 5-fold cross-validation, the framework consistently outperformed classical machine learning and deep learning baselines, attaining 99.10% accuracy, 99.11% F1-score, and 99.74% AUC, while maintaining false positives at 0.70% and false negatives at 0.20%. Efficiency profiling further showed reductions of approximately 45–55% in training time and approximately 70–80% in memory use compared with end-to-end fine-tuning, confirming its suitability for resource-constrained environments such as IoT devices, edge systems, and mobile platforms. Interpretability analyses using t-distributed stochastic neighbor embedding and gradient-weighted class activation mapping verified that the model captured semantically meaningful behavioral patterns, reinforcing trust in its operational reliability. Despite these strengths, the framework has limitations, such as the fact that adversarial robustness was not explicitly tested, leaving open the possibility that poisoning or mimicry attacks could bypass detection. In addition, the dataset size was modest, and the RL agent was limited to the tabular Q-learning.

Future research should focus on promising directions. First, enhancing adversarial robustness is a key priority. Incorporating adversarial training strategies, ensemble defenses, and RL-guided mechanisms can strengthen resilience against poisoning and mimicry attacks and ensure reliability in adversarially controlled environments. Second, scaling the evaluation to larger and more diverse datasets will be essential to validate the effectiveness of the framework against a wider spectrum of ransomware families and polymorphic variants. Third, exploring more advanced reinforcement learning algorithms may refine adaptive sample weighting, improve generalization in complex feature spaces, and further reduce the number of high-impact misclassifications.

In conclusion, TL-RL-FusionNet demonstrates how reinforcement learning can transform transfer learning into an adaptive and efficient strategy for ransomware detection. Its high accuracy while minimizing costs and enhancing interpretability establishes it as a foundation for deployable cybersecurity solutions and resilient AI-driven defenses against threats.

**CRediT authorship contribution statement**

**Jannatul Ferdous:** Conceptualization, Methodology, Python Scripting, Investigation of experimental results, Writing – original draft, and Visualization. **Rafiqul Islam**: Validation, Writing – review and editing, supervision. **Arash Mahboubi**: Validation, Writing – review & editing, Supervision. **Md Zahidul Islam:** Validation, Writing – review & editing, Supervision.

**Conflicts of interest**: The authors declare no conflicts of interest.

**Funding**: This work was supported by Charles Sturt University (CSU) PhD funding. Project Operating Fund.

**Data availability:** Data will be made available on request